\documentclass[
onecolumn]{IEEEtran}
\IEEEoverridecommandlockouts
\usepackage{cite}
\usepackage{amsmath,amssymb,amsfonts}
\usepackage{algorithmic}
\usepackage{graphicx}
\usepackage{textcomp}
\usepackage{xcolor}

\usepackage{multirow}

\def\BibTeX{{\rm B\kern-.05em{\sc i\kern-.025em b}\kern-.08em
    T\kern-.1667em\lower.7ex\hbox{E}\kern-.125emX}}
\begin{document}

\title{
Biometric-enabled Personalized 
  Augmentative and Alternative Communications
\\
\thanks{\textbf{ 
--------------------------------------------------------------------
\vspace{5mm}\\
\large This is the unpublished draft version of the paper 
{S. Yanushkevich, E. Berepiki, P. Ciunkiewicz, V. Shmerko, G. Wolbring, R. Guest, ``Biometric Technology Roadmapping for Personalized 
  Augmentative and Alternative Communication'',  that will appear in  'Computer Vision and Image Understanding'.}}}
 }

\author{\IEEEauthorblockN{\textbf{
S. Yanushkevich$^1$,
  E. Berepiki$^1$, 
	P. Ciunkiewicz$^1$,	
	V. Shmerko$^1$,  
	G. Wolbring$^{2.3}$, 
  R.  Guest$^4$}   }
\newline
\IEEEauthorblockA{
\begin{flushleft}
{\normalsize $^1$ Biometric Technologies Laboratory, Schulich School of Engineering, University of Calgary, Canada} \\ 
{\normalsize $^2$ Program in Community Rehabilitation and Disability
Studies, Cumming School of Medicine and Hotchkiss Brain Institute,
University of Calgary, Canada} \\
{\normalsize $^3$ Institute for Technology Assessment and Systems Analysis, 
Karlsruhe, Germany}\\
\normalsize $^4$ School of Electronics and Computer Science, University of Southampton, U.K.\\
}\end{flushleft}
}

\maketitle
\thispagestyle{empty}
\setcounter{page}{1}

\begin{abstract}
This study focuses on the roadmapping of biometric technologies onto personalized Augmentative and Alternative Communication (AAC), a branch of assistive technologies for people with communication disabilities. 
This technology roadmapping revolves around the proposed notions of an AAC biometric register and biometric-enabled reconfigurable AAC channels. The biometric register is referred to as a tool for acquiring and processing physiological and behavioural traits that are essential for augmentative and alternative communication. It links biometric traits, such as gestures, to intermediate traits, such as synthesized speech, for customizable communication channels. The proposed methodology is used to assess the gaps between the social and practical demands, such as assisting people with communication disabilities in the contemporary semi-automated border control, and the emerging advances in AI, such as advanced video and speech processing. We provide two case studies of the AAC that rely on hand gesture recognition and sign language word recognition, and conclude that the current accuracy of those AI technologies does not meet the practical requirements. The proposed roadmapping provides recommendations for further improvement to close these gaps.  
\end{abstract}


\tableofcontents

\section{Introduction}

Augmented and Alternative Communication (AAC)  is intrinsically a biometric-enabled technology, as it relies on capturing and transforming the biometric traits such as voice, hand and body gestures, facial expressions, etc.
The main task of AAC is to provide support to users with communication disabilities, using personalization and adaptation to the user's needs.

AAC tools are generally designed for the ``average'' communication disorder, with possibilities for adaptation to individual users. Traditional expert-driven adaptation is a costly and inefficient procedure due to the time variability of individual features. 
Personalization is a key factor in the AAC tool performance and efficiency \cite{Curtis-2022,Light-2019,McNaughton-2019}. It is implemented as an interactive process between the individual and the AAC device, aiming for step-by-step adaptation to individual features. Automating this process requires the development of specialized computational mechanisms.
The theoretical foundation of automatic adaptation addresses cognitive dynamic models \cite{Haykin-2012,Hilal-2023}, while implementation refers to self-aware computing \cite{Hoffmann-2020}. 

Despite the availability of various AAC tools, their efficacy is often compromised by individual features (e.g., physical limitations, visual disturbances, atypical speech and hand gesture patterns).

Adaptation assumes that individual features are identified in real biometric traits, while delegation means that the user intentionally (or by default) delegates preferred features into synthetic traits (virtual domain). For example, in a gesture-to-speech transformation, real video-recorded hand gestures are decoded into the synthetic audio domain, in the form of words and linguistic approximation of speech. {AAC personalization also ultimately leads to privacy issues.}

Biometrics include two kinds of traits:
\textit{Physiological} traits, such as face, fingerprints, retina, iris, and vein patterns \cite{Jain-2022}; 
\textit{Behaviour} traits \cite{Fairhurst-2017},
 such as facial expression, voice, signature, keystroke and mouse dynamics. 
The physiological biometrics are used for identity (ID) management (e.g., biometric ID and access to devices, equipment, and areas). Usability of biometric traits for the authentication of people with disabilities is studied in \cite{Brink-2019,Blanco-Gonzalo-Guest-2018}. In healthcare, they serve as biomarkers (e.g., heart rate and electroencephalogram); they are also used for identifying a human's psychological or cognitive state. 
 In contrast, non-unique behavioural traits are used for alternative identity management (e.g., gesture for smartphone ID). They are also used for  AAC.
  Synthetic traits include real and synthetic physiological traits (e.g., real face images and AI-generated face images, fingerprint images, iris images) and behaviour traits (e.g., real and synthetic voice,  facial emotions, and gestures).  Both real and synthetic biometric traits are involved in  AAC. 

{The development and deployment of biometric applications for people with disabilities are governed by various regulations.}
In particular, a Technical Report by the International Organization for Standardization (ISO) \cite{ISO-inclusive-biometric-systems} 
establishes a taxonomy of biometric ID management for individuals with disabilities. This ISO document provides recommendations on standardization in the AAC field, e.g., visual communication for audio content, user interface, sensory feedback, and alternative input methods. However, no guidance was developed regarding the design of the AAC biometric-enabled systems. 

In our study, we intend to contribute to bridging this gap.
The framework for biometric application in the AAC is proposed in our work in the form of the AAC \textit{biometric register, biometric transformations}, and the AAC \textit{interoperable technological modules}.
The biometric register is a set of biometric traits that are used in transformations for AAC purposes. 

	The concept of interoperable technological modules is an emerging trend in biometric-enabled applications, including AAC systems. This approach aims to achieve the highest levels of standardization and unification across technologies.
Interoperability offers numerous advantages, such as enhanced functionality, greater availability, improved accessibility, and increased performance. Among these, the ability to \textit{reconfigure} system components stands out as a key benefit of interoperable systems \cite{Elsahar-2019}. For instance, commonly used AAC converters, such as text-to-speech and speech-to-text systems, comprise two interoperable modules: a speech analyzer and a linguistic analyzer.
Our study aligns with this trend by proposing a systematic design methodology grounded in best practices in biometrics.

We consider AAC solutions for both individuals with disabilities and their communication partners. {Effective communication in these contexts relies on AAC systems that must satisfy interoperability criterion (the same type of AAC), compatibility criterion (different types of  AAC), and reconfiguration criterion (arrangement of AAC systems).} For example, the AAC of an individual with a hearing impairment must adapt to the AAC of an individual with a speaking impairment and that of a first responder, a healthcare worker, or a family member.
The documented references on using communication channels to reach individuals with communication disabilities include the guidance and best practice  \cite{UK_Gov_Disability_2021};  
guide to interacting with people who have disabilities \cite{DHS_IPD_2024} developed by the USA Department of Homeland Security (DHS); education in the AAC field, such as survey \cite{Sauerwein-2022} on AAC courses related to 265 speech-language pathology programs taught in the USA; and teaching about the intersectionality of disabled people using an intersectional
pedagogy framework \cite{Wolbring-2025}.
However, the communicators' teaming using their personalized AAC devices is an open problem, requiring an innovative technological solution. Our work also addresses this challenge. 

In this study, we focus on the AAC personalization that extends the recently reported results  \cite{Yanushkevich-Hawaii-2025}, where the concept of AAC channel in the context of expert elicitation was discussed. We also refer to work 
\cite{Shaposhnyk_Smart_City_2024}, where the AAC was considered as a support for first responders.


This study adopts a social model that views disability as a part of human diversity and a matter of perception,  \cite{Wright-2024}. In contrast, the medical model defines people as disabled due to their physical impairments. Our work contributes to mitigating social barriers by leveraging AAC personalization in the following ways.
1) \textbf{Taxonomized exploration:} We provide a structural exploration of the AAC field, aiming at identifying trends and challenges in biometric-enabled projections. The value of this exploration lies in creating conditions for technology transfer from biometrics and biometric-related fields to the AAC field.
2) \textbf{Novel approach to technology roadmapping:} We develop a new approach to personalized AAC technology roadmapping based on advances in AI. In this approach, existing fragmented studies are considered in the context of current demands and long-term design strategies.
3) \textbf{Creation of the AAC biometric register:} We introduce an AAC biometric register for AAC channels. Instead of the commonly accepted paradigm in the AAC research community, where biometrics are specified by default, we propose a systematic, in-depth approach that benefits from identified causal relationships between AAC demands and advanced biometric solutions, with a focus on the transformation of biometric traits.

In addition to the above contributions, other results can be useful to the developers of AAC systems. In particular, this study proposes reference protocols to support the expert elicitation process for the AAC system lifecycle. To explain why impressive achievements in the AAC field have not yet been deployed in mass-transit hubs such as airports, we conducted an experimental survey of hand gesture recognition using advanced AI tools. {We emphasize educational aspects of personalized AAC development based on fundamentals of biometric system design,  \cite{Yanushkevich-2026}.}

The paper is organized as follows. Related biometric-centric works in the field of AAC are analyzed in Section \ref{sec:audit}. A three-step technology roadmapping is introduced in Section \ref{sec:AAC-technology-landscape}. 
The results on the development of the AAC biometric register are reported in Section \ref{sec:Biometrics_register}.
In Section \ref{sec:reconfigurable-channel}, a model of the AAC channel is described. An approach to the AAC personalization is described in Sections \ref{sec:Personalization} and \ref{sec:Availability-accessibility}.
Section \ref{sec:Reference_protocol} describes the proposed usage of expert elicitation. 
Section \ref{sec:Demonstration} considers the {experimental case studies and the AAC application in automated border control.} Section \ref{sec:Discussion-conclusion-future-work} concludes this paper.

\section{Terminology}

The following terminology is used in the study:
\begin{itemize}
		\item \textit{AAC biometric register} -- a list of biometric traits used for AAC. 
				\item \textit{AAC perception-action cycle} -- key mechanism of AAC personalization based on a cognitive dynamic system model.
 \item \textit{AAC reference technology} -- advanced technology used for the AAC technology roadmapping.
\item \textit{AAC technology roadmapping} -- forecasting of the AAC technologies. 
\item \textit{Assistive biometrics,} or impairment-conditioned biometrics -- biometric traits used in AAC, such as emotion recognition conditioned by facial muscle impairments, speech or/and voice conditioned by speech/voice disabilities.
\item \textit{Biometric enabled AAC channel} -- a sequential arrangement of interoperable technologies (devices) for implementing an AAC channel for individuals with disabilities.
 	\item \textit{Biometric transformation} -- conversion of one type of biometric trait to another for the AAC, e.g., converting a hand gesture to a synthetic voice or an animated avatar, and vice versa, converting a voice to an avatar gesture.
	\item \textit{Personalization of the AAC channel} -- 1) adaptation of the AAC to individual users; 2) delegation of specific functions to the AAC.
\end{itemize}

\section{Previous works and problem formulation }\label{sec:audit}

In this section, we report the results of categorical, qualitative, and quantitative analyses of previous works, as well as comparative results. We specify the area of comparative analysis and divide the evaluation criteria into qualitative and quantitative parts.


\subsection{Specification of the comparative analysis}

The area of personalized AAC is a part of a wide range of AAC devices and systems.  Our comparative analysis addresses only this area. To be specific, we framed this particular area as follows. 
First of all, in our study and development of personalized AAC, we advocate the theory of \textit{personality computing} that concerns three problems: 1) automatic personality perception, 2) automatic personality recognition, and 3) automatic personality synthesis,  e.g., \cite{Phan-2021}. Behavioural patterns are perceived from text, audio, video, mobile phone data, digital footprints, and online games. Some of them are the object of recognition. 
Personality synthesis generates digital personalities via virtual, or embodied agents (e.g., avatars, social robots, smart homes).  

Personalization manifests itself through 1) \textit{adaptation} of an AAC device for user-specific features (e.g., a voice synthesizer to a user with speech impairment, a gesture recognizer to a user with an impaired motor function) \cite{Morrison-2023}, or/and 2) \textit{delegation} of specific features to the AAC channel (e.g., face and facial expression to avatar, voice to speech synthesizer),  \cite{Aylett-2020,Zhang-2022}.
For voice and speech synthesizers, adaptation and delegation mechanisms are implemented through changes in voice volume, tone (e.g., warm, cold, harsh, smooth), speed, pitch (highness or lowness of voice), and vocal qualities (such as breathiness, creakiness, or nasality). For hand gesture recognizers, the adaptation mechanism accounts for impairment-related biases, such as gesture speed and amplitude, as well as biases caused by finger impairments.

Personality computing includes two types of interactions: human-human and human-environment. Our study is limited by human-human communication that relies on face recognition (family members, class participants, party attenders) in conjunction with auditory perception (pronounced name) and, thus, requires personalization. Fig. \ref{fig:Two-kinds-alternative-communications} illustrates these interactions, where individual A requires two kinds of personalized support: communicating with individual B using communication messages of the AAC (our study) and in sensing the environment using {probing} messages and devices (out of our study). 

Below, we have provided a comprehensive set of markers for personalized AAC comparative analysis in the specified area in terms of criteria and categories.

\begin{figure}[!ht]
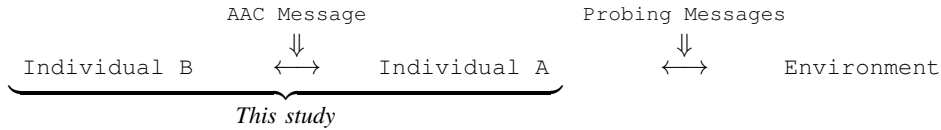

\begin{equation*}	
\underbrace{
\ensuremath{\begin{array}{c}
\texttt{\small Individual B}\\ 
\end{array}
}
\overset{
\begin{array}{c}
\texttt{\footnotesize AAC Message} \\
\Downarrow\\
\end{array}
}{\longleftrightarrow}  \hspace{-3mm}
\ensuremath{
\begin{array}{c} 
\texttt{\small Individual A}\\
\end{array}}}_{\textit{\small This study}}
\overset{
\begin{array}{c}
\texttt{\footnotesize Probing Messages} \\
\Downarrow \\
\end{array}
}{\longleftrightarrow}  \hspace{-3mm}
\texttt{\small Environment}
\end{equation*} 
\vspace{-3mm}
\caption{Specification of comparative analysis. Two kinds of AAC messages for individual A: communication with individual B (left) and sensing the environment (right). Both are in demand for personalization, but we are focused only on AAC messages.
 }\label{fig:Two-kinds-alternative-communications}
\end{figure}

\subsection{Qualitative and quantitative   comparative criteria}

To the best of our knowledge,   there is no commonly agreed set of evaluation criteria upon which to base a comparison of methodologies for the personalized AAC system design.  This is the reason to separate the qualitative and quantitative comparative analysis. 
 
We classify the AAC qualitative criteria, which are often informal and ill-defined, into the principal categories, such as taxonomy, trends, and technology roadmapping. We consolidated these explicit criteria to evaluate the personalized AAC reported in the references and standards, supplemented by the implicit criteria based on our experience of biometric-centric system design and teaching.
In contrast, the quantitative criteria are well-determined and formalized; they include datasets and various accuracy metrics that accomplish a wide range of risk assessments. Comparative summary over these criteria is given in Section \ref{sec:Demonstration}. 
{The experimental case studies helped to identify the key implementation factors and practical issues encountered by the ACC systems deployed in mass-transit security systems.}
 Specifically, using advanced technologies, we confirmed the previously reported results with potential for improvement. We identified a socio-technological gap in AAC devices in border crossing infrastructures.
 

A meaningful sample of qualitative categorized previous studies is given in Table \ref{tab:Comparative-analysis}. This sample allows researchers to make inferences about the entire AAC field. The following key categories are determined: taxonomical view and trends, societal-technology marking, personalization and expert elicitation, and education. These categories reflect various aspects of the AAC design challenges. 


\begin{table*}[!hpbt] 
\caption{A meaningful categorical sample of comparative analysis of previous and related work in the AAC design area:    taxonomy, trends, societal-technology marking, personalization and expert elicitation, and education. Aggregated categorization of this study finalizes the sample.} 
\label{tab:Comparative-analysis}
\begin{center}
\begin{small}
\begin{tabular}{l|l|l}\hline
\begin{parbox}[h]{0.15\linewidth}{\centering
\vspace{1mm} 
\textbf{Paper}
\vspace{1mm}}	
 \end{parbox}
& 
\begin{parbox}[h]{0.45\linewidth}{\centering
\vspace{1mm} 
\textbf{Description}
\vspace{1mm}}	
 \end{parbox} 
& 
\begin{parbox}[h]{0.25\linewidth}{\centering
\vspace{1mm} 
\textbf{Design Challenges}
}	
 \end{parbox}\\ \hline 
\multicolumn{3}{c}{
\definecolor{light}{gray}{.9}
\colorbox{light}{
\begin{parbox}[h]{0.9\linewidth} { \normalsize
 \textbf{\texttt{Category I: AAC Taxonomical View and Trends}}
}\end{parbox}}}
\\\hline
\begin{parbox}[h]{0.15\linewidth}{\centering \footnotesize
\vspace{1mm}
\cite{Curtis-2022} 
\vspace{1mm}}	
 \end{parbox}
& 
\begin{parbox}[h]{0.45\linewidth}{\footnotesize
\vspace{1mm} 
Systematic review and taxonomy of high-tech AAC devices and interventions using \textbf{562} articles. Focus on the dominant characteristics of high-tech AAC, research methods towards AAC design, and AAC users.
}	
 \end{parbox} 
& 
\begin{parbox}[h]{0.25\linewidth}{\footnotesize
\vspace{1mm} 
 Non-verbal communication 
(gestures, sign language, facial expression,  eye contact) and  
 group communications.
\vspace{1mm}}	
 \end{parbox}\\ \hline 
\begin{parbox}[h]{0.15\linewidth}{\centering \footnotesize
\vspace{1mm}
\cite{Blasko-2025,Light-2019,Light-2025,McNaughton-2019}
\vspace{1mm}}	
 \end{parbox}
& 
\begin{parbox}[h]{0.45\linewidth}{\footnotesize
\vspace{1mm} 
Conceptual approach to research and technology development focusing on advances in the AAC field  (accessibility, training, community support, emerging technologies). 

\vspace{1mm}}	
 \end{parbox} 
& 
\begin{parbox}[h]{0.25\linewidth}{\footnotesize
\vspace{1mm} 
 User-aware design,  personalization, accessibility, training,  optimization, precision,   practice-centric research. 
\vspace{1mm}}	
 \end{parbox}\\ \hline 
\multicolumn{3}{c}{
\definecolor{light}{gray}{.9}
\colorbox{light}{
\begin{parbox}[h]{0.9\linewidth} {\normalsize
 \textbf{\texttt{Category II: AAC Societal-Technology Marking}}
}\end{parbox}}}
\\\hline
\begin{parbox}[h]{0.15\linewidth}{\centering \footnotesize
\vspace{1mm} 
\cite{Nierling_PART_I_2018_assistive-a,Nierling_PART_II_2018_assistive-b,Nierling_PART_III_2018_assistive-c}
\vspace{1mm}}	
 \end{parbox}
& 
\begin{parbox}[h]{0.45\linewidth}{\footnotesize
Conceptual marking assistive field based on current societal and emerging technologies. 
Creation of the platform for AAC technology roadmapping.
\vspace{1mm}}	
 \end{parbox} 
& 
\begin{parbox}[h]{0.25\linewidth}{\footnotesize
\vspace{1mm}
User-aware design, autonomy, risks,  legislative and regulatory frameworks 
\vspace{1mm}}	
 \end{parbox}\\ \hline 
\multicolumn{3}{c}{
\definecolor{light}{gray}{.9}
\colorbox{light}{
\begin{parbox}[h]{0.9\linewidth} { \normalsize
 \textbf{\texttt{Category III: AAC Personalization and Expert Elicitation}}
}\end{parbox}}}
\\\hline
\begin{parbox}[h]{0.15\linewidth}{\centering \footnotesize
\vspace{1mm} 
\cite{Mordaschew2024}
\vspace{1mm}}	
 \end{parbox}
& 
\begin{parbox}[h]{0.45\linewidth}{\footnotesize
\vspace{1mm} 
There are \textbf{688} workshops for disabled workers in Germany, employing over \textbf{310,000} people at almost \textbf{2,800} locations. The research question is how to adapt an individual's personality to job requirements. 
\vspace{1mm}}	
 \end{parbox} 
& 
\begin{parbox}[h]{0.25\linewidth}{\footnotesize
\vspace{1mm} 
High potential of digital twins for disabled workers for performance maximization. 
\vspace{1mm}}	

 \end{parbox}\\ \hline 
\begin{parbox}[h]{0.15\linewidth}{\centering \footnotesize
\vspace{1mm} 
\cite{Vella-2022}
\vspace{1mm}}	
 \end{parbox}
& 
\begin{parbox}[h]{0.45\linewidth}{\footnotesize
\vspace{1mm} 
 An approach that is carried out between a team of therapists and a team of human-computer interaction researchers. The research question is how to achieve the best performance. 
\vspace{1mm}}	
 \end{parbox} 
& 
\begin{parbox}[h]{0.25\linewidth}{\footnotesize
\vspace{1mm} 
Efficient expert collaboration is a requirement for AAC personalization. 
\vspace{1mm}}	
 \end{parbox}\\ \hline 

\multicolumn{3}{c}{
\definecolor{light}{gray}{.9}
\colorbox{light}{
\begin{parbox}[h]{0.9\linewidth} {\normalsize
 \textbf{\texttt{Category IV: AAC in Education}}
}\end{parbox}}}
\\\hline
\begin{parbox}[h]{0.15\linewidth}{\centering \footnotesize
\vspace{1mm} 
\cite{Sauerwein-2022}
\vspace{1mm}}	
 \end{parbox}
& 
\begin{parbox}[h]{0.45\linewidth}{\footnotesize
\vspace{1mm} 
A survey was developed to obtain information on introductory AAC course design presently taught in the USA. 
Survey sources include \textbf{265} speech-language pathology programs, \textbf{173} AAC experts and professional educators.
\vspace{1mm}}	
 \end{parbox} 
& 
\begin{parbox}[h]{0.25\linewidth}{\footnotesize
\vspace{1mm} 
Improvement of AAC education: enrollment and course delivery, learning objectives, content, assignments, and textbooks.  
\vspace{1mm}}	
 \end{parbox}\\ \hline \hline
\multicolumn{3}{c}{
\definecolor{light}{gray}{.99}
\colorbox{light}{
\begin{parbox}[h]{0.9\linewidth} {\normalsize
 \textbf{\texttt{ Our study: Biometric Technology Roadmapping for AAC}}
}\end{parbox}}}
\\ \hline 
\begin{parbox}[h]{0.15\linewidth}{\centering \footnotesize
\vspace{1mm} 
This study
\vspace{1mm}}	
 \end{parbox}
& 
\begin{parbox}[h]{0.45\linewidth}{\footnotesize
\vspace{1mm} 
Technology roadmapping that adheres
to the best biometric practices is a step toward future AAC technologies.
\vspace{1mm}}	
 \end{parbox} 
& 
\begin{parbox}[h]{0.25\linewidth}{\footnotesize
\vspace{1mm} 
 Human-in-the-loop, self-aware computing, digital twin,
 expert elicitation.
\vspace{1mm}}	
 \end{parbox}\\ \hline 
\end{tabular}
\end{small}
\end{center}
\end{table*}

\subsubsection*{Category I: AAC taxonomical view and trends}
 
The first principal category ``AAC Taxonomical View and Trends'' in Table \ref{tab:Comparative-analysis} arose from a systematic review of 562 articles published from 1978 to 2021 \cite{Curtis-2022}, and a conceptual vision of the AAC technology landscape presented in  \cite{Blasko-2025,Light-2019,Light-2025,McNaughton-2019}. We also refer to 47 studies published between 2017 and 2025, and reviewed in \cite{Benevento-2025}. The latter review provides additional dimensions by identifying the AAC trends and design challenges on the AI platform and user-centric design. The AAC manufacturer perspective has been offered in \cite{Bonar-2024}.

\subsubsection*{Category II: AAC Societal-Technology Marking}

{The second principal Category II ``AAC Societal-Technology Marking'' in Table \ref{tab:Comparative-analysis} is formed based on the reports to the European Parliament,  \cite{Nierling_PART_I_2018_assistive-a,Nierling_PART_II_2018_assistive-b,Nierling_PART_III_2018_assistive-c,Nierling_PART_IV_2018_assistive-d}.
This category captures the conditions necessary for the AAC technology roadmapping in the socio-technological context of AAC deployment. We distinguish the three deployment scenarios of the personalized AAC system as illustrated in Fig. \ref{fig:Teachable_System}. The first scenario assumes that the AAC is implemented in the form of a software (Fig. \ref{fig:Teachable_System}$a$). This is not a cost-effective solution, as it is independent of the users' communication needs or conditions. 
The second deployment scenario relies on smartphone apps to be developed to provide access to a system placed in cloud infrastructure. This advanced approach, known as a \textit{mobile cloud computing}, allows for a drastic reduction in the cost of assistive service (Fig. \ref{fig:Teachable_System}$b$). 
Further improvements towards personalization depend on access to cloud computing resources (Fig. \ref{fig:Teachable_System}$c$). For example, an on-body communication hub for people with disabilities has been proposed in \cite{Shaposhnyk_Smart_City_2024}.
}
\begin{figure}[ht]
\begin{center}
\begin{tabular}{ccc}
\includegraphics[width=0.138\textwidth]{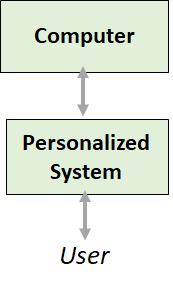}
&
~~\includegraphics[width=0.138\textwidth]{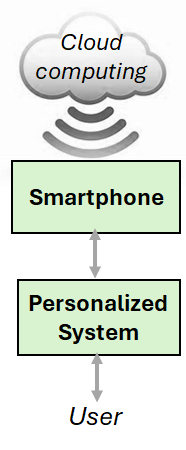}
&
~~\includegraphics[width=0.143\textwidth]{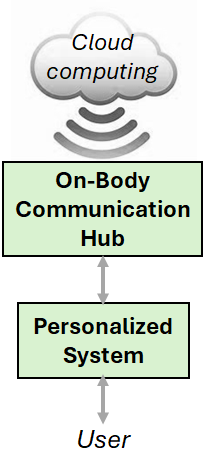}
\\
$(a)$& $(b)$& $(c)$
\end{tabular} 
\end{center}
\caption{The three scenarios for deployment of the personalized AAC systems: $(a)$ an autonomous system installed on a personal computer; $(b)$ a system that uses cloud resources via smartphone apps; $(c)$ a system that uses cloud resources via an on-body communication hub. }\label{fig:Teachable_System}
\end{figure}

\subsubsection*{Category III: AAC personalization and expert elicitation}

{The third principal Category ``AAC Personalization and Expert Elicitation''  in Table \ref{tab:Comparative-analysis} emerged from the two studies: the best practices for potential massive applications of AAC based on the digital twin concept, \cite{Mordaschew2024}, and an advanced design platform for expert elicitation, by two kinds of experts: therapists and computer engineers, \cite{Vella-2022}. The roles of the experts are of critical significance in achieving the personalization benefits.}

{Personalized AAC computing incorporates the \textit{human-in-the-loop} principle. The latter assumes that a user interacts with the AAC tool to adapt to individual features of their communication disorder \cite{Kim-2018}. This is also coherent with the concept of a digital twin. 
The \textit{digital twin} approach represents the best practice of adaptive computing \cite{Lehner-2021}. 
The idea is to adapt an AAC system to an individual user; this user must be included in a loop called a perception-action cycle, where both the AAC and the user interact. A mechanism for implementation of this principle is called a \textit{self-aware computing} \cite{Hoffmann-2020}. 
Self-awareness can be viewed as a bridge between the traditional computing systems and \textit{cognitive dynamic} model \cite{Hilal-2023}. 
The key properties of self-aware computing include the ability to learn and reason. Examples are \textit{teachable} assistive systems,  \cite{Morrison-2023}.
Fig. \ref{fig:Framework-AAC} illustrates this four-fold framework. 
}

\begin{figure}[ht]
\begin{center}
\begin{eqnarray*}
\begin{tabular}{c}
	{\texttt{\small AAC}}\\	
	{\texttt{\small Personalized}}\\	
	{\texttt{\small Computing}}\\	
	\end{tabular}
		\equiv 
\left \{
		\begin{tabular}{l}
		\texttt{ {\small Human-in-the-loop principle}} \\
		\texttt{ {\small Self-aware computing}} \\
\texttt{ {\small Cognitive dynamic model}} \\
\texttt{ {\small Digital twin approach}} \\
		\end{tabular} 
				\right .
\end{eqnarray*}
\end{center}
\caption{{ Personalized AAC computing includes a human-in-the-loop principle and its implementation using a self-aware computing, the cognitive dynamic models, and the digital twin. }}\label{fig:Framework-AAC}
\end{figure}

\subsubsection*{Category IV: AAC in education}
{The progress of the AAC field depends on the education of both the AAC designers and users. The fourth principal category, ``AAC in Education'', included in Table \ref{tab:Comparative-analysis}, is supported by the survey  \cite{Sauerwein-2022}. The survey reviewed about four hundred educational sources and suggests recommendations for advancing the area. Examples of university courses and materials that contribute to the AAC education include the following: authors  \cite{Wolbring-2025} developed the course for undergraduate and graduate level based on the intersectionality concept; the course is offered ``not only to students in critical disability studies degrees but to students in courses of any degree where disabled people are mentioned or ought to be mentioned and where the content of the degree taught is expected to impact the lived reality of disabled people.'' A textbook  \cite{Yanushkevich-2026} covers the design of the biometric-enabled systems that integrate components of the AAC for security checkpoints, mobile biometrics, forensic systems and wearables.
}

\subsection{Problem formulation and research questions}\label{sec:problem}

The above comparative exploration of the AAC field highlights several emerging problems that indicate a critical demand for technology roadmapping for personalized AAC. These problems are formulated in the form of the following research questions:

\begin{enumerate}
	\item How to apply a technology roadmapping to design the individual-centric AAC tools? What is the practical value of this methodology for the AAC field? 
    \item How to use the best practices in biometrics for AAC advances? Does technology transfer from biometrics to the AAC field benefit the latter? 
    \item Is it feasible to deploy the AAC tools in mass transit systems such as airports? What is the key obstacle? Why are the AAC tools still underutilized in applications such as airport security checkpoints?
\end{enumerate}

\section{Discovering technology milestones}\label{sec:AAC-technology-landscape}

Technology milestones traditionally represent the forecast of the AAC technology landscape. This section uses a three-step approach.

\subsection{Technology roadmapping}

In our approach, we utilize technology roadmapping for the personalized AAC field.
Technology roadmapping is a multistage strategic, long-range and long-term planning in R\&D that requires 1) monitoring the technology trends, 
2) identifying similar technologies, and 3) discovering technology opportunities. This also includes 
identification of technological gaps and obstacles, criteria of evaluation, factors of evolution, and resource allocation. 

Technology roadmapping can be understood as an ``approximation'' along the technology milestones and an ``extrapolation'', or prediction, of technology challenges. The research community has documented examples of technology roadmapping in related fields. 
The results of technology roadmapping are presented in a time-based format, such as milestones and causal graphs that link technology-related issues and business decisions, such as in \cite{Weck-2022}.

Hence, applying the technology roadmapping concept to the AAC field requires an initial state-of-the-art technology assessment, identification of trends and specifications (AAC personalization), and forecasting of the AAC technology landscape.

\subsection{Three-step approach}

The most related to our approach is \textit{causal roadmapping} developed in \cite{Dang-2023}. In this approach, the background knowledge helps to construct the causal model for reasoning on initially formulated questions. 
Building upon causal knowledge, we use different mechanisms for its extraction and exploration.
In our approach, 1) the discovery mechanism based on causal questions is replaced by the \textit{reference} technology, and 2) the causal structure is replaced by causal mapping.

Given the state-of-the-art AAC technology landscape, technology roadmapping of personalized AAC is defined as a three-step process:
\begin{enumerate}
	\item []\hspace{-9mm}\textit{Step 1:} Define {reference} technology and identify reference milestones. 
				\item []\hspace{-9mm}\textit{Step 2:} Map the reference milestones into the state-of-the-art AAC. This results in a new AAC technology landscape with clusters of milestone candidates. 
					\item []\hspace{-9mm}\textit{Step 3:} Create the AAC technology roadmap from the clustered technology landscape. 
\end{enumerate}

Therefore, the main expert decisions on road mapping concern labelling milestone candidates. Specifically, each cluster of potential milestones must be examined to roadmap the personalized AAC.

\subsection{Reference technology roadmap}

We chose the personalization model known as the digital twin model as a reference technology for roadmapping. 
Authors \cite{Thelen-2022-Part_1} proposed a five-dimensional taxonomy of digital twin: physical system, digital system, an updating engine, a prediction engine, and an optimization dimension. Rationality of this taxonomy is justified in the second part of the work  \cite{Thelen-2023-Part_2}, where the formal notions of techniques and approaches are given. 
Let us interpret each of the five taxonomical dimensions as the reference milestones and causally map into the state-of-the-art of personalized AAC tools (see Section \ref{sec:audit}).
The result of such mapping can be interpreted as the AAC technology landscape in terms of a sample of milestone candidates:

\begin{figure}[ht]
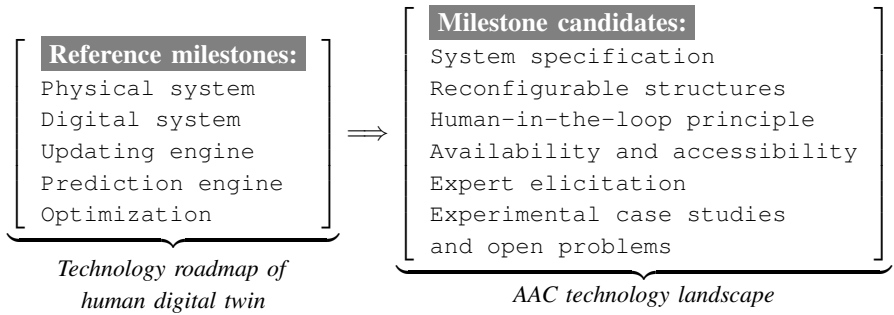

\begin{center}
\begin{eqnarray*}
\underbrace{
\left [\begin{tabular}{l}
\colorbox{gray}{\textcolor{white}{{\textbf{Reference milestones:}}}}\\	
	{\texttt{\small Physical system}}\\	
	{\texttt{\small Digital system}}\\	
	{\texttt{\small Updating engine}}\\	
	{\texttt{\small Prediction engine}}\\	
		{\texttt{\small Optimization}}\\	
	\end{tabular}
	\right  ]
	}_{
	\begin{tabular}{c}
	\textit{\small Technology roadmap of }\\
\textit{\small human digital twin}\\
		\end{tabular}
		}
\Longrightarrow		
\underbrace{
		\left [
		\begin{tabular}{l}
		\colorbox{gray}{\textcolor{white}{{\textbf{Milestone candidates:}}}}\\
		\texttt{{\small System specification}} \\
		\texttt{{\small Reconfigurable structures}} \\
\texttt{{\small Human-in-the-loop principle}} \\
\texttt{{\small Availability and accessibility}} \\
\texttt{{\small Expert elicitation}} \\
\texttt{{\small Experimental case studies}} \\
\texttt{{\small and open problems}}
		\end{tabular} \right ]
				}_{\textit{\small AAC technology landscape}}
\end{eqnarray*}
\end{center}
\caption{{Our approach is compatible with the most efficient model of personalization known as a human digital twin. The proposed technology roadmap includes the reference technology milestones as shown on the left plane.  Causal mapping of these reference milestones onto the AAC technology landscape results in a set of milestone candidates (right plane). This technology landscape is a guide to the strategy for developers of the future generation of AAC systems.  }}\label{fig:Roadmapping_AAC}
\end{figure}

Conceptually, this causal mapping attempts to translate a generalized technology model, i.e. digital twin, into another technology model. Note, this methodology of conceptual knowledge mapping is common in practice. Typically, this mapping is represented in graphical form with links between related concepts, but we use the notion of technology milestones. In our approach, the abstract technology milestones are related to their specification from the AAC field. 
For example, the first two reference milestones, $<$Physical system$>$ and $<$Digital system$>$ in roadmapping in Fig. \ref{fig:Roadmapping_AAC} are related to the potential milestone $<$System specification$>$ in the AAC technology landscape. 

\subsection{Technology milestones candidates}

The three roadmap indicators characterize the AAC technology landscape: 1) milestone candidates, 2) their state-of-the-art, and 2) their challenges. 
 Table \ref{tab:Roadstones} provides the preliminary results of technology roadmapping. 

The first milestone candidate is speculating the AAC as a biometric-enabled system. The literature reports fragmentary developments. 
Biometric technologies are necessary for personalization attributes of physical systems (acquisition, recognition, and processing of biometric traits) and digital systems (e.g., synthetic biometrics) regarding roadmapping in Fig. \ref{fig:Roadmapping_AAC}. Note that this view is congruent to the notion of human digital twin, as in \cite{Miller-2022}. Details are given in Section \ref{sec:Biometrics_register}.

\begin{table*}[!htpb]
		\caption{Challenges of milestone candidates for the AAC technology roadmapping.}
		\label{tab:Roadstones}
		\begin{center}
\begin{small}
				\begin{tabular}{c|c}
					\hline 
					\multicolumn{1}{c|}{
					\begin{parbox}[h]{0.4\linewidth} {\small \centering
\textbf{
State-of-the-art
 } 
					}\end{parbox}} &\multicolumn{1}{c}{
						\begin{parbox}[h]{0.5\linewidth}{\small \centering 
\textbf{
Challenges 
}
					}\end{parbox}}\\ \hline
\multicolumn{2}{c}{
\definecolor{light}{gray}{.9}
\colorbox{light}{
\begin{parbox}[h]{0.9\linewidth} {\vspace{1mm}
\textbf{Milestone candidate I:} \texttt{System specification} (Section \ref{sec:Biometrics_register})
\vspace{1mm}}\end{parbox}}}
\\\hline
\begin{parbox}[h]{0.4\linewidth}{\vspace{1mm} 
 Fragmentary developments
							\vspace{1mm}}
					\end{parbox} 
					&\begin{parbox}[h]{0.5\linewidth}{\vspace{1mm}
Biometric-enabled system 
							\vspace{1mm}}
					\end{parbox} \\ \hline
\multicolumn{2}{c}{
\definecolor{light}{gray}{.9}
\colorbox{light}{
\begin{parbox}[h]{0.9\linewidth} { \vspace{1mm}
\textbf{Milestone candidate II:} \texttt{Reconfigurable structures} (Section \ref{sec:reconfigurable-channel})
\vspace{1mm}}\end{parbox}}}
\\\hline
\begin{parbox}[h]{0.4\linewidth}{\vspace{1mm} 
Not specified
							\vspace{1mm}}
					\end{parbox} 
					&\begin{parbox}[h]{0.5\linewidth}{\vspace{1mm}
 AAC reconfigurable channel
							\vspace{1mm}}
					\end{parbox} \\ \hline
\multicolumn{2}{c}{
\definecolor{light}{gray}{.9}
\colorbox{light}{
\begin{parbox}[h]{0.9\linewidth} { \vspace{1mm}
\textbf{Milestone candidate III:} \texttt{Human-in-the-loop principle} (Section \ref{sec:Personalization})
\vspace{1mm}}\end{parbox}}}
\\\hline
\begin{parbox}[h]{0.4\linewidth}{\vspace{1mm} 
Teachable systems
							\vspace{1mm}}
					\end{parbox} 
					&\begin{parbox}[h]{0.5\linewidth}{\vspace{1mm}
Cognitive dynamic system model
         							\vspace{1mm}}
					\end{parbox} \\ \hline
\multicolumn{2}{c}{
\definecolor{light}{gray}{.9}
\colorbox{light}{
\begin{parbox}[h]{0.9\linewidth} {  \vspace{1mm}
\textbf{Milestone candidate IV:} \texttt{ Availability and
accessibility} (Section \ref{sec:Availability-accessibility})
\vspace{1mm}}\end{parbox}}}
\\\hline
\begin{parbox}[h]{0.4\linewidth}{\vspace{1mm} 
Fragmentary developments
							\vspace{1mm}}
					\end{parbox} 
					&\begin{parbox}[h]{0.5\linewidth}{\vspace{1mm}
Prioritized goals
							\vspace{1mm}}
					\end{parbox} \\ \hline
\multicolumn{2}{c}{
\definecolor{light}{gray}{.9}
\colorbox{light}{
\begin{parbox}[h]{0.9\linewidth} {  \vspace{1mm}
\textbf{Milestone candidate V:} \texttt{Expert elicitation} (Section \ref{sec:Reference_protocol})
\vspace{1mm}}\end{parbox}}}
\\\hline
\begin{parbox}[h]{0.4\linewidth}{\vspace{1mm} 
 Fragmentary developments
							\vspace{1mm}}
					\end{parbox} 
					&\begin{parbox}[h]{0.5\linewidth}{\vspace{1mm}
Semi-automated expert support
							\vspace{1mm}}
					\end{parbox} \\ \hline
\multicolumn{2}{c}{
\definecolor{light}{gray}{.9}
\colorbox{light}{
\begin{parbox}[h]{0.9\linewidth} {  \vspace{1mm}
\textbf{Milestone candidate VI:} \texttt{Experimental case studies} (Section \ref{sec:Demonstration})
\vspace{1mm}}\end{parbox}}}
\\\hline
\begin{parbox}[h]{0.4\linewidth}{\vspace{1mm} 
Fragmentary developments
							\vspace{1mm}}
					\end{parbox} 
					&\begin{parbox}[h]{0.5\linewidth}{\vspace{1mm}
Socio-technological best practice
							\vspace{1mm}}
					\end{parbox} \\ \hline
				\end{tabular} 
			\end{small}
		\end{center}
	\end{table*}

Challenges of the second and third milestone candidates address the core mechanism of personal system design, i.e. updating and prediction engine accordingly to the roadmapping in Fig. \ref{fig:Roadmapping_AAC}. Details of reconfigurable AAC are not systematized or specified in the AAC field. Still, developing teachable systems based on the human-in-the-loop principle in a related area, e.g., \cite{Morrison-2023}, is a step forward. Specifically, components of the AAC channel must be reconfigurable and integrated into a perception-action cycle of the cognitive dynamic model. Details are given in Sections \ref{sec:reconfigurable-channel} and \ref{sec:Personalization}.

The fourth milestone candidate represents the optimization dimension according to the reference milestone in Fig. \ref{fig:Roadmapping_AAC}, which is represented by the socio-technological dimension ``Availability and accessibility''. This is a prioritized challenge in R\&D AAC. Availability and accessibility of AAC correspond to a wide range of optimization criteria and conditions for personalized AAC design, e.g., low cost, personalized interface, computational resources (personal computer and/or cloud), on-body communication hub as an alternative (or extension) to smartphone. Details are given in Section \ref{sec:Availability-accessibility}.

Experts play a central role in the AAC personalization, as emphasized in the fifth roadstone candidate. The aim is to support the expert elicitation process. The problem is that most phases of R\&D and application of personal tools require expert knowledge from various fields. Developing semi-automated expert support and protocols for the teamwork of experts is a challenging task, as indicated in \cite{Bojke-2021}. Details are given in Section \ref{sec:Reference_protocol}.

Authors \cite{Thelen-2023-Part_2} emphasized that demonstration of a model, best practices, and open sources (e.g., free-to-use tools and datasets) for R\&D are accelerators of progress in implementing the personalization concept. The sixth milestone candidate reflects this fact. Contemporary AAC systems only partially satisfy this requirement, e.g., gesture recognition using datasets as summarized in \cite{Emporio-2025} as a part of a gesture-based communication channel. An experimental case study is needed to identify the best socio-technological practices. Details are given in Section \ref{sec:Demonstration}.

\section{System specification: AAC biometric register}\label{sec:Biometrics_register}

 This section aims to show that the AAC biometric register completely satisfies the requirements of the system specification and must be considered as a technology milestone toward personalization (first milestone candidate in Table \ref{tab:Roadstones}).

We define the AAC biometric register as a list of biometric traits for applications in the AAC. This study frames each biometric trait by the AAC's requirements and limitations. 
The identification, recognition, and application of each biometric trait are well documented. In our study, we focus on transforming biometric traits to address the AAC field.
Nine types of biometric traits are included in the AAC biometric register 
(Fig. \ref{fig:Biometric_Register}):

\begin{figure*}[!ht]
\begin{center}
\includegraphics[width=0.65\textwidth]{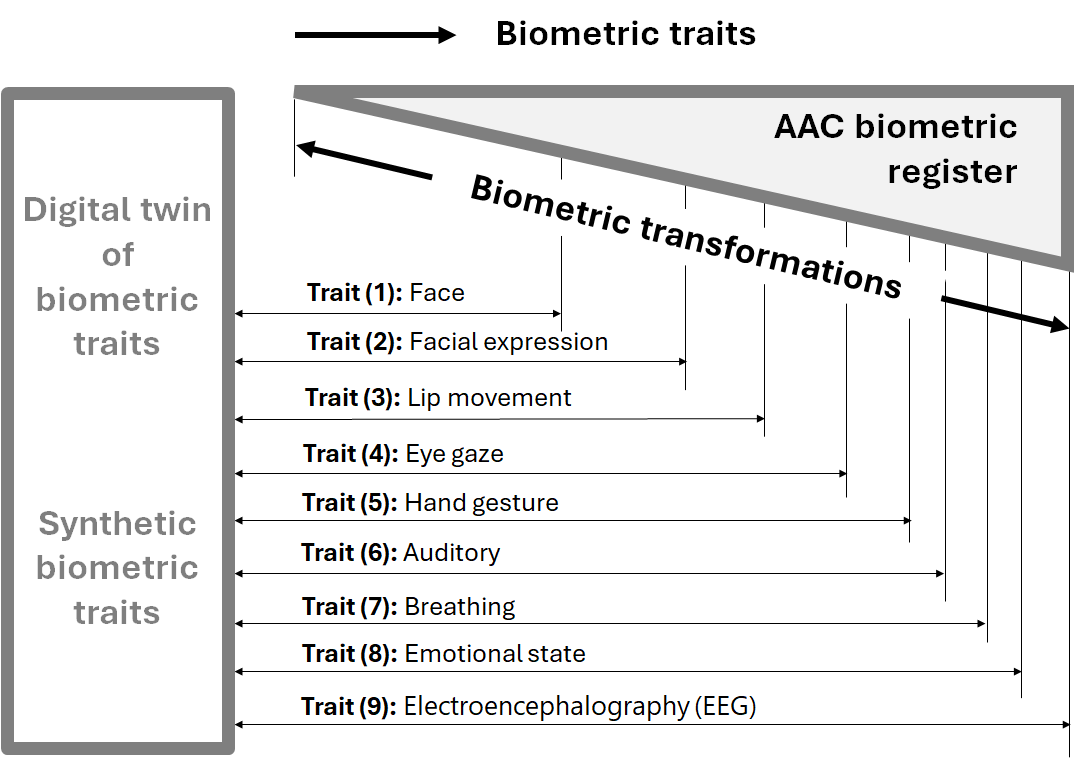}
\caption{{Illustration of the biometric-centric AAC personalization within the dimensions of biometric traits, biometric transformations, and their digital twins, including synthetic biometrics. The  AAC biometric register that captures the 9 most usable biometric traits is the core of the personalized AAC technology roadmapping. The relationships between the biometric traits are determined by the transformations, or translations, to enable communication for individuals with disabilities.  
}}\label{fig:Biometric_Register}
 \end{center}
\end{figure*}

 \begin{itemize}
	\item []\hspace{-7mm}\textit{Trait of type (1)}: Face is considered as a transmitter of information encoded as emotion state, lip movement, eye gaze, and head movement. 
	A synthetic face can be used to represent themselves, express their identity in online communication. It is a common practice of avatar visualization with integrated voice and emotions.
	\item []\hspace{-7mm}\textit{Trait of type (2)}: Facial expressions convey information about the emotional states. Given a facial image, facial expressions can be automatically recognized, synthesized, and converted to other traits (e.g., gesture, voice, and lip movements).
	Synthetic facial expressions can be integrated into avatar face visualization and manipulated intentionally or synchronized with real emotions.
	\item []\hspace{-7mm}\textit{Trait of type (3)}: Lip movement can be decoded and converted to other traits (e.g., facial expressions, gestures, and auditory signals).
	Synthetic lip movement can be integrated into a talking avatar-visualized face, converted to text, or converted to avatar-visualized gestures. 
	\item []\hspace{-7mm}\textit{Trait of type (4)}: Eye gaze can be decoded and converted to other traits (e.g., gestures, auditory signals, lip movements, and facial expressions).
	Synthetic eye gaze is a useful data structure for modelling and training people to use these alternative communication methods.
\item []\hspace{-7mm}\textit{Trait of type (5)}: Hand gesture can be recognized, synthesized, and converted to other traits (e.g., facial expressions, lip movements, auditory signals, and emotional states).
	Synthetic hand gestures can be used for conversions such as text-to-gestures and speech-to-gesture, teaching, training, and modelling. 
\item []\hspace{-7mm}\textit{Trait of type (6)}: Auditory signals (words and speech) can be recognized, and converted to facial expressions, lip movements, and eye gaze.
	Synthetic voice and speech can be used for various transformations of data structures, e.g., gesture-to-speech, text-to-speech, gaze-to-words, and emotion-to-words.
\item []\hspace{-7mm}\textit{Trait of type (7)}: Breathing encoded signals can be decoded and converted to other traits that are useful for communication, e.g., auditory signals, facial expressions, and gestures.
	Synthetic breathing can be used for teaching and training, and modelling with the avatar technique assistance.
\item []\hspace{-7mm}\textit{Trait of type (8)}: Emotional state (e.g., stress, anxiety, and depression) can be recognized using a set of physiological signals such as heart rate variability, blood pressure, breathing function, EEG, electromyography,
photoplethysmogram, respiratory activity, and peripheral skin temperature. These states can be transformed into avatar visualized synthetic traits such as facial expression, lip movements, and eye gaze. 
\item []\hspace{-7mm}\textit{Trait of type (9)}: EEG is the base of the brain-to-brain interface for mutually exchanging decoded neural information between two brains. Recent achievements, e.g., brain-speech interface referred to  \cite{Metzger-2023,Moses-2021}, demonstrated potential to be useful for the AAC. 
	Synthetic EEGs are useful for training and modelling.
\end{itemize}

\section{Reconfigurable structures: AAC channel}\label{sec:reconfigurable-channel}

Adaptation of the AAC to the individual user (or group of people) requires a reconfiguration (structural or parametric) of the system (second milestone candidate in Table \ref{tab:Roadstones}).

\subsection{Basic definitions}

Communication is a joint activity aiming at transmitting and receiving information. The message is sent through the channel to another person who has to understand it. 
\begin{itemize}
	\item Human communication abilities include verbal communication (message exchange in linguistic form) and non-verbal communication (message exchange using body language, touch, and facial expressions).
	\item Models of communication are understood as abstract and simplified representations of the communication process. Typical components of the human communication model are the person-sender, the message, the channel, the person-receiver, and the feedback loop (Fig. \ref{fig:Model-AAC}).
	\item Notion of channel addresses the senses for perceiving the message: seeing, hearing, touching, smelling, and tasting. 
	\item Models of human communications are based on face-to-face interactions with integrating verbal and nonverbal communication, reflecting personality (dialect, styles and manner of talking and facial expressions, tone of voice, arms and hands).
	\item Alternative communications are represented by the scheme Face (gesture)-to-AAC to Face (gesture). 
\end{itemize}

Models of communication are categorized according to applications. The AAC require the development of specialized models where components of channels can imitate speech, hearing, and vision using recognition techniques and synthetic equivalents, e.g., synthetic voice, synthetic facial expressions, synthetic language such as speech and text/written constructions, as well as hand gesture languages. These and other components must be connected to achieve the desired goal, e.g., conversion of written words to voice, transforming speech to gestures using avatar visualization. 

\begin{figure}[!ht]
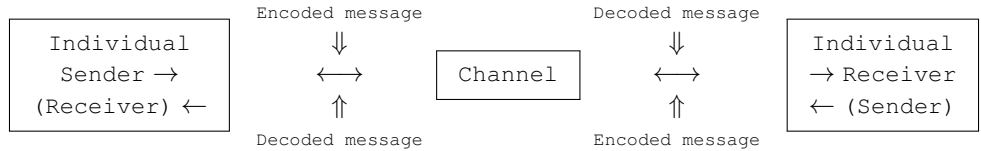

\begin{equation*}
\fbox{\ensuremath{
\begin{array}{c}
\texttt{\small Individual}\\ 
\texttt{\small Sender} \rightarrow\\
\texttt{\small (Receiver)}\leftarrow\\
\end{array}
}}
\enskip
\begin{array}{c}
\texttt{\scriptsize Encoded message}\\
\Downarrow\\
\longleftrightarrow\\
\Uparrow\\
\texttt{\scriptsize Decoded message}\\
\end{array} 											
\fbox{\ensuremath{
\begin{array}{c}
\texttt{\small Channel}
\end{array}
}}
\begin{array}{c}
\texttt{\scriptsize Decoded message}\\
\Downarrow\\
\longleftrightarrow\\
\Uparrow\\
\texttt{\scriptsize Encoded message}\\
\end{array}
\enskip
\fbox{\ensuremath{
\begin{array}{c}
\texttt{\small Individual}\\
\rightarrow	\texttt{\small Receiver}\\
\leftarrow\texttt{\small (Sender)}\\
\end{array}
}}
\end{equation*} 
\vspace{-3mm}
\caption{{An AAC model of the communication process between two individuals: the channel is designed to perceive the message through seeing, hearing, touching, smelling, and tasting; messages (e.g., words, facial expressions, gestures) are encoded and transmitted via the channel from one individual to another, and decoded as required; the automated recognition tools help both the encoding and decoding. For example, decoding could be the process of converting a written text into a spoken language, while encoding could be converting a spoken language into a written text. 
}}\label{fig:Model-AAC}
\end{figure}

\subsection{Taxonomy of the AAC channel}


The biometric register in Fig. \ref{fig:Biometric_Register} includes suggestions to distinguish original and synthetic biometric traits. Let us apply this taxonomical view to the AAC components.
Denote:
\begin{itemize}
\item [$-$]Biometric trait as \texttt{B-trait} -- behavioural biometrics; this may include facial biometrics, body gesture, and impairment-conditioned behavioural biometrics, including facial expressions, eye gaze, eye blinking, lip movements.
\item [$-$]Intermediate data structure as \texttt{I-trait} structure -- additional non-biometric elements incorporated into the AAC process, such as synthetic biometrics, text, and avatar visual and/or audio representations.		
\end{itemize}

Two basic components, the B-trait and the I-trait, can encode an arbitrary message from a person with communication disabilities. Details of transformations are illustrated in Fig.\ref{fig:transformations}.

\begin{figure}[ht]
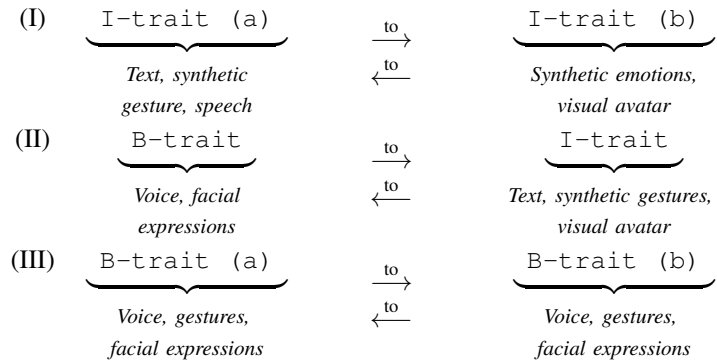

\begin{center}
\begin{tabular}{cccc}
(I) &
$\underbrace{
\begin{array}{c}
\texttt{I-trait (a)}\\
\end{array}}_{\begin{array}{c}
\textit{\footnotesize Text, synthetic}\\
\textit{\footnotesize gesture, speech}\\
\end{array}}
\hspace{5mm}
$&$
\begin{array}{c}
\stackrel{\text{to}}{\longrightarrow}\\
\stackrel{\text{to}}{\longleftarrow}\vspace{-10mm}
\end{array}
$&$
\hspace{5mm}
\underbrace{
\begin{array}{c}
\texttt{I-trait (b)} \\
\end{array}}_{
\begin{array}{c}
\textit{\footnotesize Synthetic emotions,}\\
\textit{\footnotesize visual avatar}\\
\end{array}
}$\\
(II)&
$\underbrace{
\begin{array}{c}
\texttt{B-trait}\\
\end{array}}_{\begin{array}{c}
\textit{\footnotesize Voice, facial}\\
\textit{\footnotesize expressions}\\
\end{array}}
\hspace{5mm}
$&$
\begin{array}{c}
\stackrel{\text{to}}{\longrightarrow}\\
\stackrel{\text{to}}{\longleftarrow}\vspace{-10mm}
\end{array}
$&$
\hspace{5mm}
\underbrace{
\begin{array}{c}
\texttt{I-trait} \\
\end{array}}_{
\begin{array}{c}
\textit{\footnotesize Text, synthetic gestures, }\\
\textit{\footnotesize visual avatar}\\
\end{array}
}$
\\
(III)&
$\underbrace{
\begin{array}{c}
\texttt{B-trait (a)}\\
\end{array}}_{\begin{array}{c}
\textit{\footnotesize Voice, gestures, }\\
\textit{\footnotesize facial expressions}\\
\end{array}}
\hspace{5mm}
$&$
\begin{array}{c}
\stackrel{\text{to}}{\longrightarrow}\\
\stackrel{\text{to}}{\longleftarrow}\vspace{-10mm}
\end{array}
$&$
\hspace{5mm}
\underbrace{
\begin{array}{c}
\texttt{B-trait (b)} \\
\end{array}}_{
\begin{array}{c}
\textit{\footnotesize Voice, gestures, }\\
\textit{\footnotesize facial expressions}\\
\end{array}
}$
\end{tabular}
\end{center}
\caption{{The three modes of transformation of biometric traits in a personalized AAC: (I) within the synthetic domain; (II) between the real and synthetic domains, and (III) within the real domain.}}\label{fig:transformations}
\end{figure}

{The first transformation mode in Fig.\ref{fig:transformations}(I)  
represents the conversion of an I-trait of type (a) into an I-trait of type (b), and vice versa. For example, in \cite{Hyppa-2023}, synthetic emotions are converted into a synthetic voice. The second mode of transformation in Fig. \ref{fig:transformations}(II) illustrates the relationship between the real and synthetic biometric traits. The third mode of transformation in Fig. \ref{fig:transformations}(III)  reflects the conversion between the real biometric traits. These modes reflect various possibilities for composing an AAC channel for a given individual. For example, the individual's emotional state can be detected in voice pitch, which is synchronized with both the facial expression and body language, as well as with the biomarkers such as heart rate variability and blood pressure. Emotion visualization involves all three modes; for example, facial expressions and hand or body gestures are synthesized in response to the text and speech words such as feeling good (bad, happy, sad), smiling, being confused, angry, disgusted or surprised.
}

\subsection{Examples of biometric-enabled AAC channels}

Consider the AAC channel for communication with a hearing-impaired traveller. The ABC provides the AAC channels as drawn in Fig. \ref{fig:AAC_deaf-mute_individual}. Linguistic approximation is needed between recognized gestures to construct a \texttt{Speech} data structure. For this purpose, the mediated data structure \texttt{Text} is needed. In response, gestures must be extracted from speech. Text data structure is preferable for this processing. Conversion Gesture-Speech can be used for arbitrary sign language (e.g., American, Indian, Filipino, etc.) for people who are deaf or speech-impaired. Note that this AAC Traveller-Officer interaction can be significantly simplified in the case of Traveller-Machine interactions because gesture recognition can be embedded in automated border control.

\begin{figure*}[!htp]
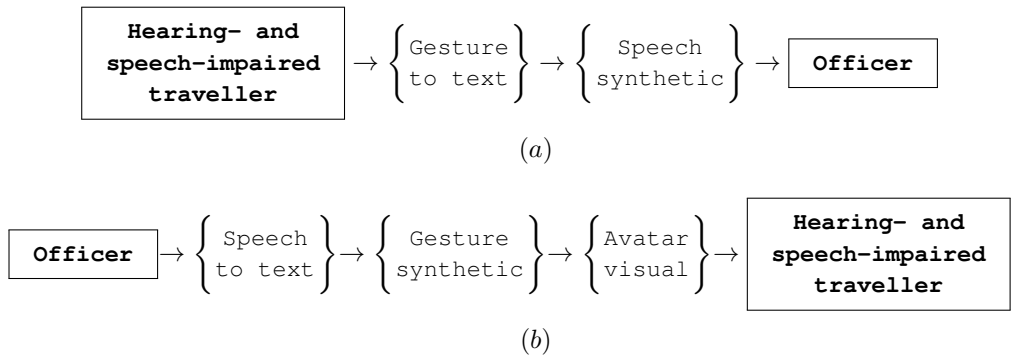

	\centering
\begin{eqnarray*}\label{scheme:}
\fbox{\begin{tabular}{c} 
			\texttt{\small \textbf{Hearing- and }}\\
		\texttt{\small \textbf{speech-impaired}}\\
				\texttt{\small \textbf{traveller}}\\
		\end{tabular}}
\rightarrow
\Biggl\{{\hspace{-1.9mm}
\begin{tabular}{c} 
		\texttt{\small Gesture}\\
				\texttt{\small to text}\\
		\end{tabular}}\hspace{-1.9mm}\Biggr\}
\rightarrow
		\Biggl\{{\hspace{-1.9mm}
\begin{tabular}{c} 
		\texttt{\small Speech}\\
						\texttt{\small synthetic}\\
		\end{tabular}}\hspace{-1.9mm}\Biggr\}
\rightarrow
\fbox{\begin{tabular}{c} 
		\texttt{\small \textbf{Officer}}\\
		\end{tabular}}
\end{eqnarray*}			
	\begin{center}
$(a)$
\end{center}	
\begin{eqnarray*}			
\fbox{\begin{tabular}{c} 
		\texttt{\small \textbf{Officer}}\\
		\end{tabular}}
\hspace{-0.9mm} \rightarrow
\Biggl\{{\hspace{-1.9mm}
\begin{tabular}{c} 
		\texttt{\small Speech}\\
				\texttt{\small to text}\\
		\end{tabular}}\hspace{-1.9mm}\Biggr\}
\hspace{-0.9mm} \rightarrow
		\Biggl\{{\hspace{-1.9mm}
\begin{tabular}{c} 
		\texttt{\small Gesture}\\
						\texttt{\small synthetic}\\
		\end{tabular}}\hspace{-1.9mm}\Biggr\}
\hspace{-0.9mm} \rightarrow
		\Biggl\{{\hspace{-1.9mm}
\begin{tabular}{c} 
		\texttt{\small Avatar}\\
				\texttt{\small visual}\\
		\end{tabular}}\hspace{-1.9mm}\Biggr\}
\hspace{-0.9mm} \rightarrow 
\fbox{\begin{tabular}{c} 
		\texttt{\small \textbf{Hearing- and }}\\
		\texttt{\small \textbf{speech-impaired}}\\
				\texttt{\small \textbf{traveller}}\\
		\end{tabular}}
\end{eqnarray*}	
	\begin{center}
$(b)$
\end{center}	
\caption{The AAC for hearing- and speech-impaired individuals and officers: $(a)$ gesture biometric is transformed to speech biometric using a text as intermediate data structure; $(b)$ speech biometric is transformed to gesture biometric that is visualized using an avatar technology.
}\label{fig:AAC_deaf-mute_individual}
\end{figure*}

\subsection{Generalization of AAC reconfiguration}

The AAC channel for two persons can be generalized for teamwork. Fig. \ref{fig:Teamwork-six-individuals}$(a)$ introduces a teamwork of two groups of people with varying communication disabilities, Team I and Team II. This teamwork is possible only if the configurable AAC devices, {called a Communication Hub}, are available. For example, the communication path in Fig. \ref{fig:Teamwork-six-individuals}$(b)$ states that first individual (Teammate-1) with communication disability AAC$_1$ communicate with sixth individual (Teammate-6) using reconfiguration ${AAC}_{1\rightarrow 6 }$. An example of a multi-target personalization has been introduced in \cite{Mordaschew2024}. The authors studied a virtual representation of disabled workers (individual impairments, skills, and dynamic behaviour) for prospective production planning and production control.

\begin{figure*}[!h]
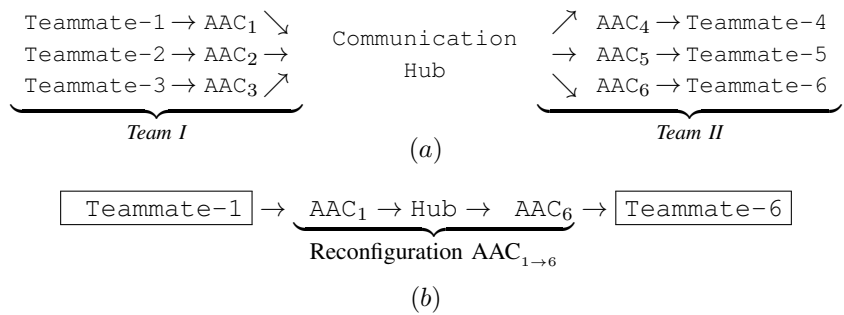

	\begin{center}
	\begin{parbox}[h]{0.9\linewidth} {
\begin{equation*}
	\underbrace{
\begin{array}{ccc} 
\texttt{\small Teammate-1}
&\hspace{-3mm}\rightarrow  \texttt{\small AAC}_1 
&\hspace{-3mm}\searrow  \\
\texttt{\small Teammate-2}
&\hspace{-3mm}\rightarrow  \texttt{\small AAC}_2 
&\hspace{-3mm}\rightarrow \\
\texttt{\small Teammate-3}
&\hspace{-3mm}\rightarrow \texttt{\small AAC}_3 
&\hspace{-3mm}\nearrow \\
	\end{array}
	}_{\textit{\footnotesize Team I}}
	\mbox{
	\begin{tabular}{c} 
	\texttt{\small Communication}\\
	\texttt{\small Hub}\\
		\end{tabular}}
			\underbrace{
	\begin{array}{clcc} 
	 \nearrow&\hspace{-1mm}\texttt{\small AAC}_4 &\hspace{-3mm}\rightarrow
	&\hspace{-3mm} \texttt{\small Teammate-4}
\\
	\rightarrow&\hspace{-1mm}\texttt{\small AAC}_5 &\hspace{-3mm}\rightarrow
	&\hspace{-3mm} \texttt{\small Teammate-5}\\
	\searrow &\hspace{-1mm}\texttt{\small AAC}_6 &\hspace{-3mm}\rightarrow
	&\hspace{-3mm} \texttt{\small Teammate-6}
	\end{array}
	}_{\textit{\footnotesize Team II}}
	\end{equation*}
}\end{parbox}\\
\vspace{-3mm}
$(a)$\\
\vspace{-3mm}
$$\fbox{\texttt{ Teammate-1}} \rightarrow 
\underbrace{
\texttt{ AAC}_1 \rightarrow 
\texttt{Hub} \rightarrow 	
\texttt{ AAC}_6 
}_{\text{\small Reconfiguration AAC}_{1\rightarrow 6 }}
\rightarrow 
\fbox{\texttt{Teammate-6}}$$\\

$(b)$
\caption{Team work using a generalized AAC reconfiguration: $(a)$ Communication between two teams with varying communication disabilities. Any teammate of the first team with personalized AAC 1,2, and 3 can communicate with any teammate of the second team with personalized AAC 4, 5, and 6; {$(b)$ example of communication between the first and sixth individuals (Teammate-1 and Teammate-6).} }
		\label{fig:Teamwork-six-individuals}
		\end{center}
	\end{figure*}

\section{Human-in-the-loop: Personalization}\label{sec:Personalization}

{AAC personalization can be achieved using an adaptation strategy. There are three main levels of adaptation within the personalized AAC systems: a user-configured adaptation (manual settings), a context-aware adaptation (semi-automated), and a dynamic AI-driven adjustment (automated, learned from interactions with the user in real time). Adaptation is implemented using the perception-action model of computing with human-in-the-loop, also known as self-aware computing.}
This corresponds to
the 3rd and 4th milestone candidates displayed in Table \ref{tab:Roadstones}).
 
In Fig. \ref{fig:perception-action-model}, the perception-action cycle includes a Perceptor that learns individual features and performs reasoning; an Actuator that illuminates the individual patterns and acts to adapt a device to these patterns; and an Individual that can evaluate an adaptation using stimuli (control signs from the actuator) to report the current adaptation state to the perceptor using appropriate sensors. The cycle is repeated to achieve an acceptable level of adaptation.

\begin{figure}[!ht]
\begin{center}
\begin{tabular}{c}
\includegraphics[width=0.4\textwidth]{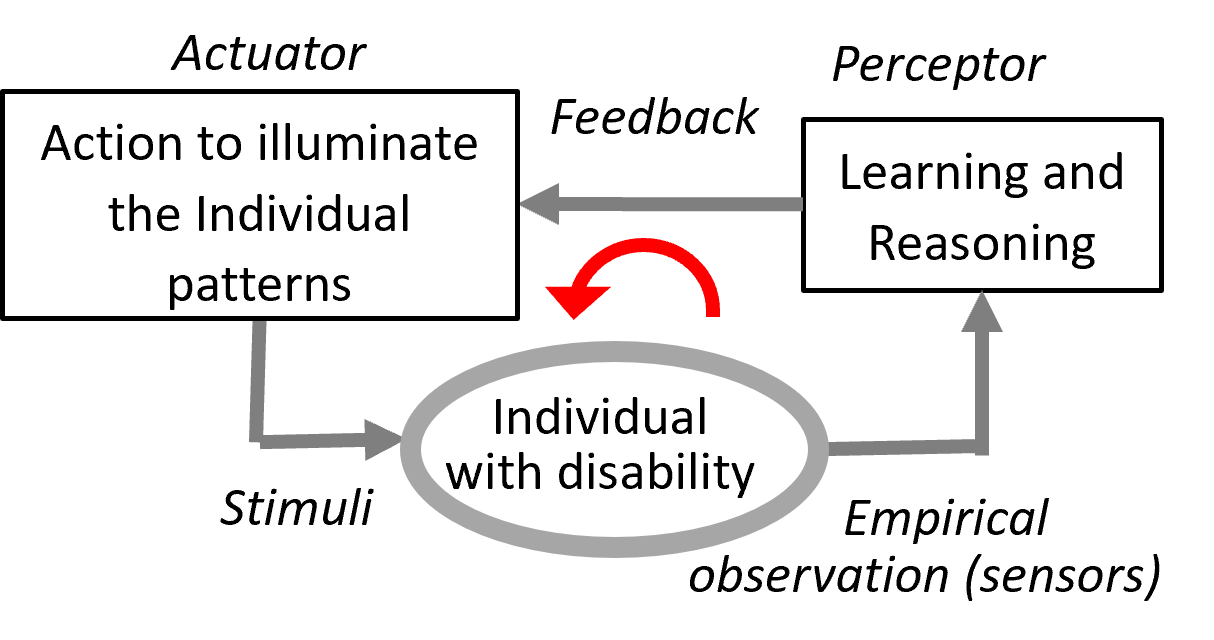}
\end{tabular}
 \caption{{The key design principle of the personalized AAC is the human-in-the-loop, which means a perception-action cycle with two functional parts: Perceptor that perceives an individual, and Actuator, that ``controls'' the perceptor by acting within the AAC system.
}
}\label{fig:perception-action-model}
 \end{center}
\end{figure}

The perception-action cycle is the core of a contemporary system, e.g., speech sound treatment such as   \cite{Benway-2024}.
 Work \cite{Benedictis-2023} used a cognitive dynamic model (attention function, memory function, perceptual function and
higher-level cognitive function) for robotic personalized assistive service for people with disabilities, while the Classification of Functioning, Disability and Health (ICF) model was proposed in \cite{WHO-ICF-2024}.
{Thus, we witness the emergence of a new discipline in education, -- cognitive AAC systems. This discipline will provide fundamentals for the design, development, and applications of the next generation of personalized AAC tools. This approach is closely related to the idea of a cognitive system first proposed in \cite{Haykin-2012}. The authors argue that a system is ``cognitive'' if it performs the following four functions: the perception-action cycle, memory, attention, and intelligence.}

\section{Availability and accessibility}\label{sec:Availability-accessibility}

AAC availability and accessibility address a wide spectrum of societal and technological parameters and characteristics, including cost-related, user satisfaction, user geography, personalization, and mobile communication hub between user and resources (smartphone or its alternative as an on-body communication device). Advanced AAC technologies can be prohibitively expensive and may not be covered by insurance or healthcare systems. This limits access to many people with communication disabilities. Distributed AAC resources can radically decrease the cost.

In \cite{Shaposhnyk_Smart_City_2024}, a concept of an on-body e-hub for people with disabilities was adopted from first responders. The tasks of the on-body mobile computing tools are delegated to the city infrastructure and cloud platform, leaving only energy-saving pre-processing computing. Miniature energy-saving, portable, comfortable, and low-cost on-body sensing and communication tools are embodied into a belt, wristband, or jacket, e.g., \cite{Lampea-2018}.

The on-body hub helps to explore AAC resources using e-health, a multidimensional sustainability system that includes technology, organization, economic, social, and resource assessments.

\section{Expert elicitation}\label{sec:Reference_protocol}

Identification, specification, and optimization of the AAC tools for an individual is a complicated task that, typically, addresses a group of experts (fifth milestone candidate in Table \ref{tab:Roadstones}). Teamwork of each expert requires framing for various criteria such as field of expertise, individual and joint functions, and responsibilities. Protocols and supported mechanisms are commonly accepted regulators of teamwork.

\subsection{Challenges}

Demands on optimization of AAC are reviewed in \cite{Vogel-2024}. This work was motivated by the complex needs of multi-system neurodegenerative diseases, such as ataxia, which is characterized by variability in time. These people have multiple difficulties with communication, including hearing, vision, and writing/typing. Moreover, these people require regular reassessment of AAC and tailoring of AAC strategies, which addresses adaptive AAC.

Expert tasks include identifying utterances from atypical speech to choose the speech understanding tools and corresponding word meaning predictor \cite{Venugopalan-2023}. Note that predicting a text and phrase selection is also needed to minimize the amount of typing required. For example, eye-tracking AAC devices can be used with text-to-speech technology to verbalize the message. A group of experts from various fields must frame communication and computing functions. This collective knowledge provides the key parameters for AAC configuration. 

{\cite{Vella-2022} called the expert elicitation process a ``co-design'', in which the two teams of experts are involved in the user-centric AAC design: the therapists and the human-computer interaction researchers.  Such expert elicitation was implemented using the two components: 1) an Editor (responsible for morphology and contents of the AAC), and 2) a Generator of an executable AAC with a configuration interface. }

\subsection{Reference protocol for expert elicitation}

The reference protocol manifests in various forms, such as computer templates or benchmarks for best practices and reporting, serving as ``standard tools that allow valuating and comparing different systems or components according to specific characteristics such as performance, dependability, and security''. The work  \cite{Bojke-2021} proposes a set of requirements (principles) for effective elicitation. Based on this work, we emphasize the following requirements to experts: Transparency (ensuring that the process is open and understandable); Usefulness (must be relevant and applicable to the specific decision-making context, such as people with disabilities); and Adaptivity (allowing for adjustments based on the expertise and skills of the individuals involved). These principles aim to enhance the effectiveness and reliability of the reference protocol in evaluating and comparing different systems and components. A joint User-in-the-loop and Expert-in-the-loop paradigm for high-intensity, evidence-based speech sound treatment was recently developed in \cite{Benway-2024}.

 \subsection{Example of reference protocol}

An example of a reference protocol is given in
Fig. \ref{fig:Reference_Protocol}. It supports an expert judgment, and is represented by the two marginal conclusions (\textbf{X} and \textbf{Y} dimensions). Dimension \textbf{X} includes the available emerging AAC technologies. Sample includes a text-to-speech (gesture) and related technologies, personalization techniques, tactile-to-voice conversions, techniques for prediction (writing or speaking words and gestures), emotion-to-text (voice) technologies, and avatar-related techniques. Dimension \textbf{Y} is constructed accordingly to the  Classification of Functioning, Disability and Health (ICF) developed in \cite{WHO-ICF-2024}. Work  \cite{Benedictis-2023} provides an approach on how the ICF model can be used for robot personalization of assistive behaviour for people with disabilities. In this work, researchers interpreted a subset of ICF stable and dynamic variables to determine what services a particular user needs.
 Guidance for this protocol is as follows:

\begin{figure*}[!ht]
\begin{center}
\includegraphics[width=0.69\textwidth]{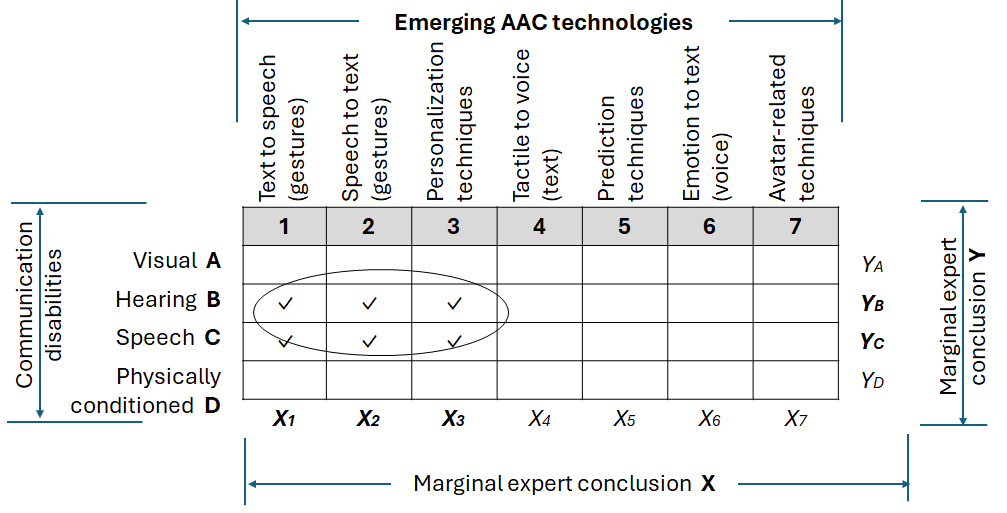}
\caption{Example of reference protocol for expert elicitation of the AAC field. Each cell represents an expert judgment on the recommended assistive tools. The clusters represent collective expert judgments; the transparency of the judgments is visualized using the dimensions of communication disability type and emerging AAC technologies.
}\label{fig:Reference_Protocol}
 \end{center}
\end{figure*}

\begin{itemize}
	\item Each cell represents the type of communication disability with respect to its mitigation possibilities, represented by available technologies.
		\item Expert task is to group the cells in clusters, aiming to cover the communication disabilities (often multiple and condition specified) by multiple technological opportunities.
		\item These clusters must be optimized with respect to cost, interoperability, availability, and accessibility. 
		\item Marginal expert conclusion \textbf{X} addresses the summary of technologies for a given individual. This conclusion is necessary for preliminary cost estimation, interoperability specification, and assessment of availability and accessibility.
		\item Marginal expert conclusion \textbf{X} addresses the communication disability landscape that can vary in some diseases and conditions. 
\end{itemize}
 
Experts must often deal with hundreds of such cells. The essence of this approach is the transparency of the judgments and the possibility of processing in a semi-automated manner, that is, both in manual and automation modes. 	
For example, cluster $\{B1,B2,B3,C1,C2,C3\}$ represents scenarios of security control in mass-transit hubs, where text-to-speech and speech-to-text technologies, and their personalization are employed to address both hearing and speech impairments.

\section{Experimental case studies and open problems}\label{sec:Demonstration}

The sixth milestone candidate for AAC technology roadmapping, shown in Table \ref{tab:Roadstones} addresses two aspects: an experimental demonstration of the AAC achievements and trends, and an audit of open problems in the AAC applications. 
In this section, we report two case studies: gesture and sign language word recognition, and open problems within the contemporary semi-automated border control.  .

\subsection{Experimental case study I: Gesture recognition}\label{sec:}

The experimental study aims to explore the practical aspects of integrating gesture recognition into automated border control (Task 3 of the problem formulation, subsection \ref{sec:problem}). 
{A sample of a comparative quantitative analysis is given in Table \ref{tab:Comparative-analysis-Gesture-Sign_Lan}; it lists the authors of the reported results (the first column), the names of datasets (the second column), the utilized computing platform (the third column), and the achieved recognition accuracy (the fourth column).
} 
The above advances were used in our experiments, building on the previous results reported in \cite{Lai-2020}.

\begin{table*}[!hpbt] 
\caption{Comparative quantitative analysis of previous work on gesture recognition and  American sign language recognition (sample). } 
\label{tab:Comparative-analysis-Gesture-Sign_Lan}
\begin{center}
\begin{small}
\begin{tabular}{l|l|l|l}\hline
\begin{parbox}[h]{0.12\linewidth}{\centering
\vspace{1mm} 
\textbf{Paper}
\vspace{1mm}
}	
 \end{parbox}
& 
\begin{parbox}[h]{0.19\linewidth}{\centering
\vspace{1mm} 
\textbf{Dataset}
\vspace{1mm}}	
 \end{parbox}
&
\begin{parbox}[h]{0.35\linewidth}{\centering
\vspace{1mm} 
\textbf{Computing Platform}
\vspace{1mm}}	
 \end{parbox} 
& 
\begin{parbox}[h]{0.2\linewidth}{\centering
\vspace{1mm} 
\textbf{Accuracy}
}	
 \end{parbox}\\ \hline 
\multicolumn{4}{c}{
\definecolor{light}{gray}{.9}
\colorbox{light}{
\begin{parbox}[h]{0.9\linewidth} {\centering \normalsize
 \textbf{\texttt{Gesture recognition}}
}\end{parbox}}}
\\\hline
\begin{parbox}[h]{0.11\linewidth}{\centering \footnotesize
\vspace{1mm}
\cite{Dang-2022}
\vspace{1mm}}	
 \end{parbox}
& 
\begin{parbox}[h]{0.19\linewidth}{\footnotesize
\vspace{1mm} 
 Static HAnd PosE (SHAPE) Dataset
\vspace{1mm}}	
 \end{parbox} 
& 
\begin{parbox}[h]{0.35\linewidth}{\footnotesize
\vspace{1mm} 
 Deep neural network models.
\vspace{1mm}}	
 \end{parbox} 
& 
\begin{parbox}[h]{0.2\linewidth}{\footnotesize
\vspace{1mm} 
Highest reported accuracy of \textbf{96}\%.
\vspace{1mm}}	
 \end{parbox}\\ \hline 
\begin{parbox}[h]{0.11\linewidth}{\centering \footnotesize
\vspace{1mm}
\cite{Shi-2019}
\vspace{1mm}}	
 \end{parbox}
& 
\begin{parbox}[h]{0.19\linewidth}{\footnotesize
\vspace{1mm} 
 NTU-RGBD and  \\
 Kinetics-Skeleton Dataset
\vspace{1mm}}	
 \end{parbox} 
& 
\begin{parbox}[h]{0.35\linewidth}{\footnotesize
\vspace{1mm} 
Two-stream adaptive graph convolutional network for skeleton-based on action recognition.
\vspace{1mm}}	
 \end{parbox} 
& 
\begin{parbox}[h]{0.2\linewidth}{\footnotesize
\vspace{1mm} 
Highest reported accuracy \textbf{95}\% and \textbf{58}\%, respectively.
\vspace{1mm}}	
 \end{parbox}\\ \hline 
\begin{parbox}[h]{0.11\linewidth}{\centering \footnotesize
\vspace{1mm}
\cite{Lai-2020} 
\vspace{1mm}}	
 \end{parbox}
& 
\begin{parbox}[h]{0.19\linewidth}{\footnotesize
\vspace{1mm} 
 DHG-14/28, Dataset
 \cite{De-Smedt-2016}
\vspace{1mm}}	
 \end{parbox} 
& 
\begin{parbox}[h]{0.35\linewidth}{\footnotesize
\vspace{1mm} 
Ordered-neuron long-short-term-memory based on recurrent neural networks.
\vspace{1mm}}	
 \end{parbox} 
& 
\begin{parbox}[h]{0.2\linewidth}{\footnotesize
\vspace{1mm} 
Highest reported accuracy of \textbf{85}\%. 
\vspace{1mm}}	
 \end{parbox}\\ \hline 
\multicolumn{4}{c}{
\definecolor{light}{gray}{.9}
\colorbox{light}{
\begin{parbox}[h]{0.9\linewidth} {\centering \normalsize
 \textbf{\texttt{Sign language  recognition}}
}\end{parbox}}}
\\\hline
\begin{parbox}[h]{0.11\linewidth}{\centering \footnotesize
\vspace{1mm}
\cite{Miah-2024} 
\vspace{1mm}}	
 \end{parbox}
& 
\begin{parbox}[h]{0.19\linewidth}{\footnotesize
\vspace{1mm} 
WLASL 100,
300, and
2000,
Dataset \cite{Li-2020}
\vspace{1mm}}	
 \end{parbox} 
& 
\begin{parbox}[h]{0.35\linewidth}{\footnotesize
\vspace{1mm} 
A two-stream multistage graph convolutional model with attention and residual connections designed to extract spatiotemporal contextual information.
\vspace{1mm}}	
 \end{parbox} 
& 
\begin{parbox}[h]{0.2\linewidth}{\footnotesize
\vspace{1mm} 
\textbf{90}\%, \textbf{69}\%, and \textbf{63}\% highest reported accuracy for each subset, respectively.
\vspace{1mm}}	
 \end{parbox}\\ \hline 
\begin{parbox}[h]{0.11\linewidth}{\centering \footnotesize
\vspace{1mm}
\cite{amorim2019stgcn} 
\vspace{1mm}}	
 \end{parbox}
& 
\begin{parbox}[h]{0.19\linewidth}{\footnotesize
\vspace{1mm} 
ASLLVD Dataset
\vspace{1mm}}	
 \end{parbox} 
& 
\begin{parbox}[h]{0.35\linewidth}{\footnotesize
\vspace{1mm} 
Spatiotemporal graph convolutional network
 using human skeleton data. 
\vspace{1mm}}	
 \end{parbox} 
& 
\begin{parbox}[h]{0.2\linewidth}{\footnotesize
\vspace{1mm} 
 Highest accuracy of \textbf{61}\%.
\vspace{1mm}}	
 \end{parbox}\\ \hline 
\begin{parbox}[h]{0.11\linewidth}{\centering \footnotesize
\vspace{1mm}
\cite{Li-2020}
\vspace{1mm}}	
 \end{parbox}
& 
\begin{parbox}[h]{0.19\linewidth}{\footnotesize
\vspace{1mm} 
WLASL 100,
300,
1000,
2000,
Dataset \cite{Li-2020}
\vspace{1mm}}	
 \end{parbox} 
& 
\begin{parbox}[h]{0.35\linewidth}{\footnotesize
\vspace{1mm} 
Pose-based temporal graph convolution networks and comparison with an inflated 3D convolutional network.
\vspace{1mm}}	
 \end{parbox} 
& 
\begin{parbox}[h]{0.2\linewidth}{\footnotesize
\vspace{1mm} 
Highest reported accuracy of \textbf{64}\%. 
\vspace{1mm}}	
 \end{parbox}\\ \hline 
\begin{parbox}[h]{0.11\linewidth}{\centering \footnotesize
\vspace{1mm}
\cite{Hu-2023}
\vspace{1mm}}	
 \end{parbox}
& 
\begin{parbox}[h]{0.19\linewidth}{\footnotesize
\vspace{1mm} 
HANDS17, MSASL, WLASL, and SLR500 datasets
\vspace{1mm}}	
 \end{parbox} 
& 
\begin{parbox}[h]{0.35\linewidth}{\footnotesize
\vspace{1mm} 
Self-supervised pre-trainable framework with model-aware hand prior incorporated.
\vspace{1mm}}	
 \end{parbox} 
& 
\begin{parbox}[h]{0.2\linewidth}{\footnotesize
\vspace{1mm} 
\textbf{95}\%, \textbf{83}\%, \textbf{81}\%, and \textbf{95}\% highest reported accuracies, respectively.
\vspace{1mm}}	
 \end{parbox}\\ \hline 
\begin{parbox}[h]{0.11\linewidth}{\centering \footnotesize
\vspace{1mm}
\cite{Yu-2025}
\vspace{1mm}}	
 \end{parbox}
& 
\begin{parbox}[h]{0.19\linewidth}{\footnotesize
\vspace{1mm} 
How2Gesture Dataset
\vspace{1mm}}	
 \end{parbox} 
& 
\begin{parbox}[h]{0.35\linewidth}{\footnotesize
\vspace{1mm} 
A Transformer-based 3D sign-language generation model using diffusion denoising, enhanced text conditioning, and staged joint-level action synthesis.
\vspace{1mm}}	
 \end{parbox} 
& 
\begin{parbox}[h]{0.2\linewidth}{\footnotesize
\vspace{1mm} 
Highest reported Joint Mean Absolute Error of \textbf{0.079}.
\vspace{1mm}}	
 \end{parbox}\\ \hline \hline

\multicolumn{4}{c}{
\definecolor{light}{gray}{.99}
\colorbox{light}{
\begin{parbox}[h]{0.9\linewidth} {\centering \normalsize
 \textbf{\texttt{Our Study: Biometric Technology Roadmapping for AAC}}
}\end{parbox}}}
\\\hline \hline 
\begin{parbox}[h]{0.11\linewidth}{\centering \footnotesize
\vspace{1mm} 
This study
\vspace{1mm}}	
 \end{parbox}
& 
\begin{parbox}[h]{0.19\linewidth}{\footnotesize
\vspace{1mm} 
WLASL-2000, Dataset
\cite{Li-2020}
\vspace{1mm}}	
 \end{parbox} 
& 
\begin{parbox}[h]{0.35\linewidth}{\footnotesize
\vspace{1mm} 
 Light-weight transformer-based model, trained on 2D/3D keypoints.
\vspace{1mm}}	
 \end{parbox} 
& 
\begin{parbox}[h]{0.2\linewidth}{\footnotesize
\vspace{1mm} 
Highest reported accuracy of \textbf{69}\%. 
\vspace{1mm}}	
 \end{parbox}\\ \hline
\end{tabular}
\end{small}
\end{center}
\end{table*}

For our experiment  reported in \cite{Lai-2020}, we used the Dynamic Hand Gesture-14/28 DHG-14/28 dataset \cite{De-Smedt-2016}. This dataset comprises gestures performed by twenty unique individuals, each completing five iterations of 14 gestures (such as grab, tap, expand, pinch, rotations, swipes, and shake) using two types of finger configurations, resulting in 28 sets of gestures and a total of 2800 sequences. The depth information is stored as images with a $480 \times 640$ resolution at 16 bits. The skeleton data contains 22 joint locations of a hand, described in both 2D and 3D coordinates. An ensemble model, consisting of four classifiers, was employed, along with knowledge distillation techniques, including ``dark'' knowledge, which refers to hidden information within the models. Further details of the experiments reported in \cite{Lai-2020} are provided in the Supplementary material.

Despite utilizing advanced resources, the achieved accuracy was 86\% using skeleton joint positions, and 85\% using fusion of depth and skeleton information for various hand gestures. Notably, some gestures were recognized with even higher accuracy, such as the Shake gesture, which reached 95\%, while the lowest recognition rates were approximately 70\%. The results reported in the literature at that time were comparable to our case study reported in \cite{Lai-2020}; for example, researchers \cite{Koh-2019} achieved 82\%.

The achieved accuracy can be interpreted in the practice of AAC communication as follows. There is a $15\%-30\%$ risk that 1) Conversation between the user who communicates using gestures and the first responder can be misunderstood or incorrectly understood; and 2) Service time can be increased because of needs, multiple clarifications, and mistaken decisions. 3) In the security applications, such as border control, the accuracy of biometric trait transformation can be intentionally used for an attack. 

In particular, the risk of $15\%-30\%$ indicates that the gesture is not recognized, which can lead to miscommunication or misunderstanding, and can be interpreted as an unintentional \textit{semantic attack}; this can also be used intentionally for a created attack, or deception. Semantic attacks are directed against a decision-making process. A gesture recognition inaccuracy provides a wide range of opportunities for manipulating the traveller's information in conversation with a border control officer. 

Countermeasures to semantic attacks include increasing the accuracy of gesture recognition.

\subsection{Experimental case study II: Sign language word  recognition}

Computational approaches for translating gestures into spoken or written language are essential for developing AAC technologies that serve diverse Sign Languages (SLs) worldwide. Real-time SL recognition requires interpreting dynamic gestures, yet many existing approaches employ high-level models with substantial computational costs and processing latencies. These limitations pose significant challenges for real-time applications, particularly emergency response scenarios where first responders require immediate communication with impaired individuals, as in \cite{Shaposhnyk_Smart_City_2024}, and for deployment on embedded on-body systems \cite{Yanushkevich-Hawaii-2025}.

Recent studies demonstrate that less complex models, including Convolutional Neural Networks \cite{garcia2016real} and algorithms such as DyFAV (Dynamic Feature Selection and Voting)  \cite{paudyal2019comparison}, achieve satisfactory performance in alphabet-based American SL (ASL) recognition. 
{A sample of comparative analysis is given in Table \ref{tab:Comparative-analysis-Gesture-Sign_Lan}. }
Note that higher accuracies have been reported for other sign languages,  particularly at word and sentence levels. 
However, ASL recognition exhibits lower accuracy due to limitations including insufficient large-scale datasets, class imbalance, and restricted vocabulary coverage. Primary ASL datasets (ASLLVD, How2Sign, and WLASL) provide substantial data volumes, yet high-performing models typically utilize subsets with limited vocabulary sizes that do not represent real-world scenarios.

Our experimental study, reported in \cite{Berepiki-2025}, identified efficient strategies to reduce model complexity through lightweight transformer encoders with simplified inputs without sacrificing accuracy. We investigated keypoint-only approaches for ASL recognition, experimented with single-modality and multimodal keypoint inputs, and applied modern architectures with hyperparameter optimization to enhance performance. This approach enables efficient parameter evaluation while establishing foundations for transfer learning in the specialized ASL domain.

\begin{figure}[!ht]
\begin{center}
\includegraphics[width=0.98\textwidth]{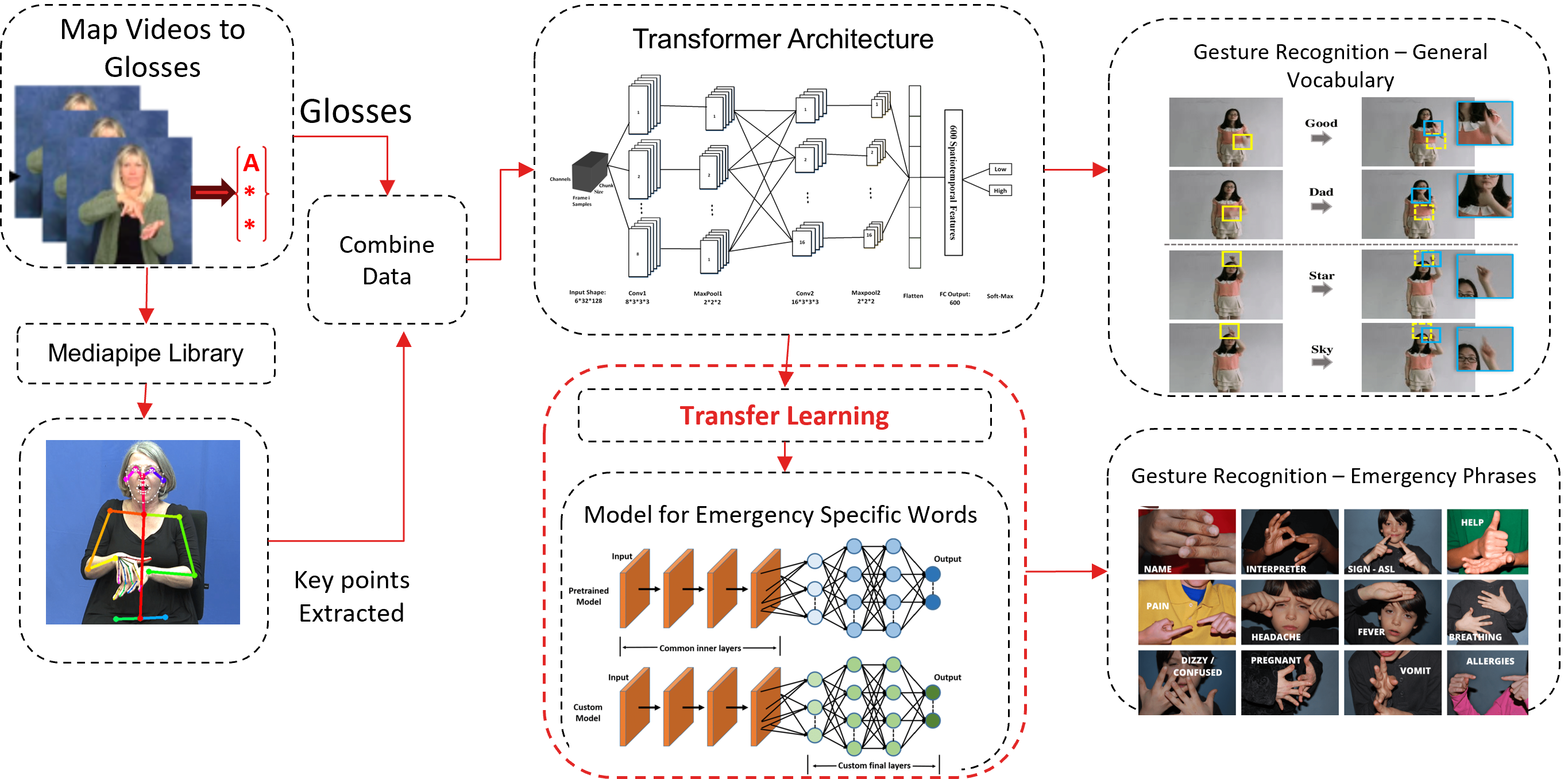} 
\caption{Illustrations of the experimental setup for recognizing a subset of American Sign Language (ASL) as reported in \cite{Berepiki-2025}. Pose key points are extracted from the video of individuals performing the ASL glosses (gestures for phrases), and become the input for a deep network, such as a transformer; transfer learning is performed to train the glosses recognition model on a subset of phrases used in emergencies. 
}\label{fig:Exper}
 \end{center}
\end{figure}

Fig. \ref{fig:Exper} illustrates an experimental setup reported in \cite{Berepiki-2025} for the recognition of a subset of body gestures for ASL that could be used by the emergency department to communicate with individuals who are hard of hearing and/or speaking. To recognize word-level ASL glosses from video data, in \cite{Berepiki-2025}, we developed a pipeline that models temporal pose dynamics through keypoint information, eliminating appearance-based features. We considered two architectures.
Fig. \ref{fig:short}a presents the first architecture, a unimodal design utilizing hand and body keypoints to evaluate performance with minimal biometric features. To assess whether additional modalities enhance recognition, we introduced a second architecture (Fig. \ref{fig:short}b) incorporating multimodal input: hand, body, and face mesh keypoints.

In the multimodal configuration, face and hand keypoints are processed separately. For example, the MediaPipe  \cite{mediapipe2024} provides over four times more facial keypoints than combined hand and body keypoints. To prevent overfitting to the denser facial representation and underutilization of hand and body data, each input type is independently projected to a weighted embedding space before fusion, ensuring balanced representation across modalities during training.

\begin{figure*}[ht]
  \centering
    \includegraphics[width=0.95\textwidth]{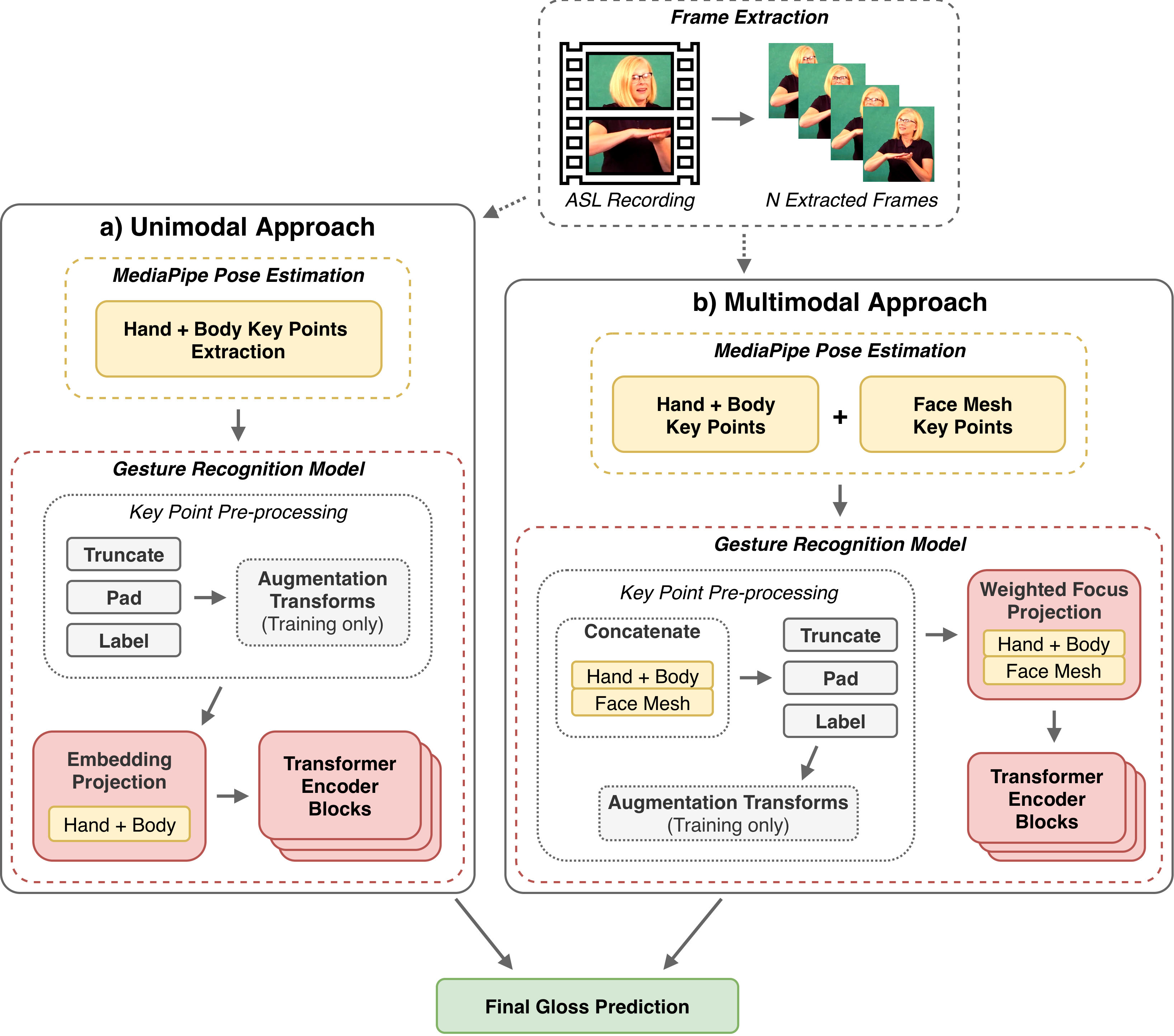}
  \caption{Comparison of ASL recognition pipelines utilizing distinct keypoint modalities: (a) a unimodal approach employing hand and body keypoints extracted via MediaPipe, and (b) a multimodal strategy incorporating hand, body, and face mesh keypoints with a weighted focus projection module. Both pipelines comprise frame extraction, keypoint estimation, preprocessing, and transformer-based gloss prediction; the multimodal configuration leverages additional facial cues to enhance modelling.}
  \label{fig:short}
\end{figure*}


Top-1, Top-5, and Top-10 accuracy metrics are reported in Table \ref{tab:wlasl_results}, categorized by model input modality: pose-based, appearance/3D-based, and keypoint-based variants integrating hand, body, and face mesh features.

Pose-TGCN marginally outperformed Pose-GRU (Top-1: 23.65\%, Top-5: 51.75\%, Top-10: 62.24\%) in the pose-based category, suggesting temporal graph modelling enhances pose representation, albeit with limited overall accuracy relative to other methods. I3D achieved the highest accuracy (Top-1: 32.48\%, Top-5: 57.31\%, Top-10: 66.31\%) in the appearance-based category, highlighting the effectiveness of spatiotemporal features from video. However, this gain incurs substantial computational and memory overhead. VGG-GRU reported lower performance across all benchmarks, highlighting the role of temporal dependencies in sign language recognition.

\begin{table}[ht]
\caption{Model performance comparison trained on WLASL-2000 for Top-1, Top-5, and Top-10 accuracy (\%) reported in \cite{Berepiki-2025}.}
  \label{tab:wlasl_results}
  \centering
    \begin{tabular}{llccc}
     \hline
      \textbf{Type} & \textbf{Model} & \textbf{Top-1} & \textbf{Top-5} & \textbf{Top-10} \\
    \hline 
      \multirow{2}{*}{Pose-Based} 
        & Pose-GRU \cite{Li-2020}   & 22.54 & 49.81 & 61.38 \\
        & Pose-TGCN \cite{Li-2020}  & 23.65 & 51.75 & 62.24 \\
     \hline 
      \multirow{2}{*}{Appearance/3D-Based} 
        & I3D \cite{Li-2020}       & 32.48 & 57.31 & 66.31 \\
        & VGG-GRU \cite{Li-2020}   & 8.44  & 23.58 & 32.58 \\
     \hline 
      Hands + Body & Transformer \\Keypoints 
        &  encoder (ours) & 22.96& 53.12 & 65.60 \\
      \hline 
      Hands + Body & Transformer  \\ + Face Mesh Keypoints 
        & encoder (ours) & 21.00 & 51.30 & 62.70 \\ 
      \hline 
      Hands + Body + Face\\ Mesh with Focus & Transformer \\
      Projection, Keypoints
        & encoder (ours) & \textbf{26.06} &  \textbf{56.91} & \textbf{68.97} \\
    \hline 
    \end{tabular}
  \end{table}

The proposed Transformer-Encoder model with exclusively hand and body keypoints achieved a Top-10 accuracy of 65.60\%, approaching that of I3D without requiring full-frame video. This demonstrates the efficacy of keypoint-only representations when combined with attention mechanisms and appropriate training strategies.

Adding face mesh keypoints to hand and body keypoints resulted in a performance decrease (Top-1: 21.00\%, Top-5: 51.30\%, Top-10: 62.70\%), suggesting that the model was splitting its efforts across too many inputs. Extending this multimodal approach with focus projection resulted in better generalization (Top-1: \textbf{26.06\%}, Top-5: \textbf{56.91\%}, Top-10: \textbf{68.97\%}). This result indicates that dense visual data is not strictly required to achieve strong model performance. While appearance-based models exhibit stronger Top-1 accuracy, keypoint-based approaches offer strong performance metrics with significantly reduced computational complexity.


{Summarizing, the proposed model attained improved accuracy of 69\% for the gesture recognition using the  'Top-10' ranking approach, compared to state-of-the-art alternatives where  66\% was achieved for 'Top-10' strategy, demonstrating competitive performance alongside computational efficiency. This model, framed by the WLASL-2000 dataset, has the potential for real-world AAC applications, such as emergency and low-resource contexts, e.g., on-body cameras for first responders \cite{Shaposhnyk_Smart_City_2024}. It should be noted that the achieved accuracy does not satisfy the requirements for high-risk applications such as security checkpoints.  
}



\subsection{Open problems in the AAC application for semi-automated border control}

The technology gaps in servicing travellers with disabilities must be addressed by integrating the AAC tools into mass-transit systems such as airports and seaports. This integration encounters multiple socio-technological barriers reported in \cite{Oostveen-2018,Nierling_PART_II_2018_assistive-b}. In particular, the experimental case study, shown in the previous subsection, shows that the accuracy of hand gesture recognition and ASL gesture recognition from videos in a controlled experiment is not satisfactory. Deployment of algorithms developed on laboratory-collected data to real-life scenarios generally leads to a significant degradation in accuracy. In the context of border control, this creates further problems for traveller throughput and overall system security.

\begin{figure*}[!!ht]
\begin{center}
\includegraphics[width=0.95\textwidth]{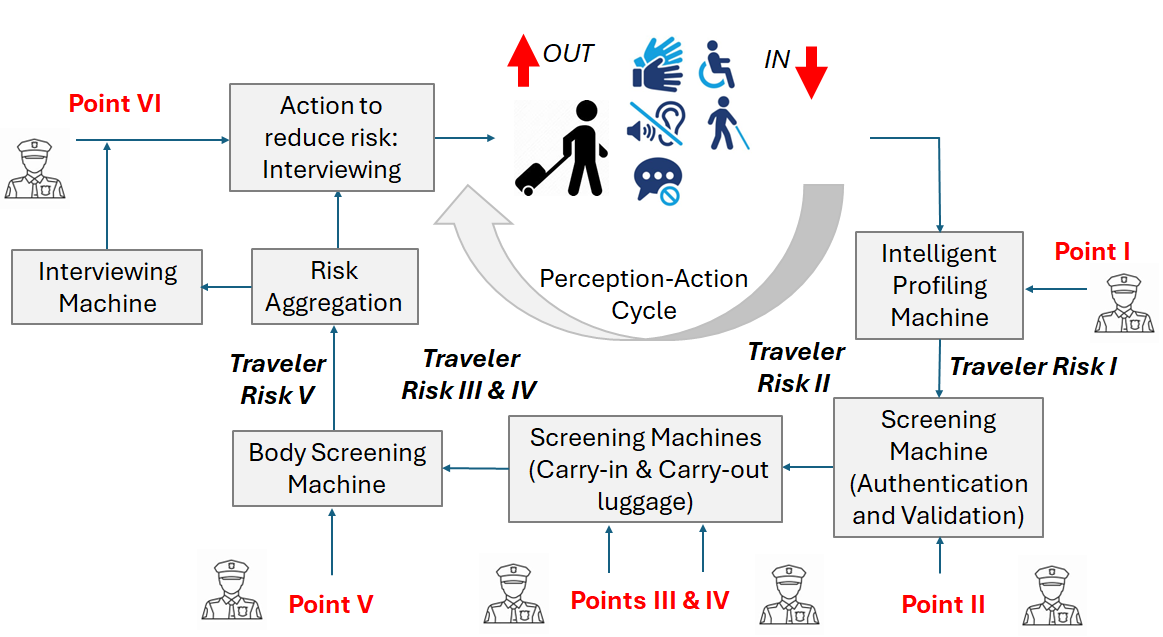}
\end{center}
\caption{{Example of open problems in the AAC applications for mass-transit hubs. Six security points of the semi-automated border control where the personalized AAC tools are required for communication between the system operator and the traveller. Each communication session using AAC aims at assessing the traveller's risk. Accuracy and errors of the AAC  system can increase traveller risk at security points, resulting in the alarm level.
}}\label{fig:AAC_Embedding}
\end{figure*}

Fig. \ref{fig:AAC_Embedding} represents an architecture of a contemporary semi-automated border control system \cite{Jain-2022}. There are six points of communication and traveller risk assessment. Each of these points requires the deployment of AAC. Point I, Intelligent Profiling Machine, requires preliminary information about the traveller, such as the type of communication disability and e-health provider, for Risk I assessment. Point II, Screening Machine (Authentication and Validation), involves standard actions on automatic authentication using e-ID for Risk II assessment. In particular cases, an alternative semi-automated gate may be used. Points III and IV, such as the screening machine (Carry-in and carry-out luggage), require communication regarding standard security protocols for risk III and IV assessments. Point V, Body Screening Machine, requires communication on specific security actions for Risk V assessment. In Point VI, Interviewing Machine, the AAC tools should provide a wide range of understandable interactions between the traveller and the officer to reduce traveller risk. These six points of traveller risk assessment are integrated into a risk perception-action cycle. For example, if a traveller does not satisfy the risk regulations, the process can be repeated, including the acquisition of additional information about the traveller.



\section{Summary and conclusions}\label{sec:Discussion-conclusion-future-work}

This study aims to update the socio-technological pathways of a social model that considers disability as a part of human diversity and a matter of perception. Personalized alternative communication solutions are central to this social model. Our study focuses on the technology-centric systematization of the AAC field, identification of trends and obstacles in AAC system deployment, and assessment of potential future perspectives.
The technology roadmapping methodology has proven suitable for these purposes. Thanks to the accessibility of cloud-based distributed resources and computational intelligence tools, complex AAC models with embedded mechanisms for automated adaptation to users and environments are now available in practice.

Our study focuses on the automatic personalization of AAC devices, a problem characterized by high computational complexity. However, recent technological advancements provide unique opportunities for implementing and deploying personalized devices, which tend to reduce costs and improve accessibility and availability.
Our study leads to several conclusions and recommendations:
1) The developed AAC technology roadmapping is a useful methodology for combining achievements and advances toward the technology horizon. In our approach, reference milestones of advanced technology (such as digital twins) are mapped onto the AAC field, highlighting achievements and gaps. This conclusion is based on an analysis of the effects of this mapping.
2)  Utilizing best practices from biometrics is beneficial for the AAC field. This includes a systematic view of resources, such as an AAC biometric register and biometric-enabled, reconfigurable AAC channels.
3)  The current accuracy of hand gesture recognition {and sign language word recognition} in AAC does not meet the requirements for automated security control in mass-transportation hubs; therefore, these AAC systems cannot yet be deployed in such settings. {Our work contributes to bridging this technological gap.}

{Adaptation is a strategic reserve for further improvements of the gesture-aware personalized AAC. This part of the biometric technology roadmapping is based on the assumption that users' private data are available for training the AAC intelligent tools. It enters the area of privacy preservation, such as differential privacy, an advanced technology for mitigating the risk of unintended data disclosure. This area is being intensively developed, e.g., in \cite{NIST-Near-2025}.}

\subsection*{Acknowledgment}
This work was supported in part by the Social Sciences and Humanities Research Council of Canada (SSHRC) through the Grant NFRF-2021-00277 ''Emergency Management Cycle-Centric R\&D: From National Prototyping to Global Implementation''.

\end{document}